\documentclass[aps,twocolumn,groupedaddress,showpacs,nofootinbib]{revtex4-1}
\usepackage{graphicx}
\usepackage{amsmath}
\usepackage{multirow}
\usepackage{hyperref}
\usepackage[usenames]{color}
\usepackage{url}
\hypersetup{
    colorlinks=true,
    linkcolor=red,
    citecolor=red,
}
\def\be{\begin{equation}}
\def\ee{\end{equation}}
\def\bea{\begin{eqnarray}}
\def\eea{\end{eqnarray}}

\begin{document}

\title{Searching for scalar gravitational interactions in current and future cosmological data}

\author{Alireza Hojjati$^{1,2}$, Aaron Plahn$^{2}$, Alex Zucca$^{2}$, Levon Pogosian$^{2}$, Philippe Brax$^{3}$, Anne-Christine Davis$^{4}$, Gong-Bo Zhao$^{5,6}$}
\affiliation{$^1$Department of Physics and Astronomy, University of British Columbia, Vancouver, BC, V6T 1Z1, Canada}
\affiliation{$^2$Department of Physics, Simon Fraser University, Burnaby, BC, V5A 1S6, Canada}
\affiliation{$^3$Institut de Physique Theorique, CEA, IPhT, CNRS, URA 2306, F-91191Gif/Yvette Cedex, France}
\affiliation{$^4$DAMTP, Centre for Mathematical Sciences, University of Cambridge, Wilberforce Road, Cambridge CB3 0WA, U.K.}
\affiliation{$^{5}$National Astronomy Observatories, Chinese Academy of Science, Beijing, 100012, People's Republic of China}
\affiliation{$^{6}$Institute of Cosmology and Gravitation, University of Portsmouth, Portsmouth, PO1 3FX, United Kingdom}

\begin{abstract}
Modified gravity theories often contain a scalar field of gravitational strength which interacts with matter.
We examine constraints on the range and the coupling strength of a scalar gravitational degree of freedom using a subset of current data that can be safely analyzed within the linear perturbation theory. Using a model-independent implementation of scalar-tensor theories in {\tt MGCAMB} in terms of two functions of the scale factor describing the mass and the coupling of the scalar degree of freedom, we derive constraints on the $f(R)$, generalized chameleon, Symmetron and Dilaton models. Since most of the large scale structure data available today is from relatively low redshifts, only a limited range of observed scales is in the linear regime, leading to relatively weak constraints. We then perform a forecast for a future large scale structure survey, such as \emph{Large Synoptic Survey Telescope} (LSST), which will map a significant volume at higher redshifts, and show that it will produce much stronger constraints on scalar interactions in specific models. We also perform a principal component analysis and find that future surveys should be able to provide tight constraints on several eigenmodes of the scalar mass evolution.
\end{abstract}

\maketitle

\section{Introduction}

A non-vanishing cosmological constant, $\Lambda$, is the simplest and the most common explanation of the observed cosmic acceleration \cite{Riess:1998May,Perlmuter:1999Dec}. Because, gravitationally, $\Lambda$ is equivalent to the large vacuum energy predicted in particle physics, its value requires a technically unnatural fine-tuning 
\cite{Weinberg:1988cp,Burgess:2013ara} in order to be consistent with observations. The cosmological constant could be embedded in a larger class of dark energy models, where dynamics dictate the value of the vacuum energy. Because of the absence of apparent violation of Lorentz invariance in the Universe, dark energy is commonly described by the field theory of a scalar. Usually, some degree of fine-tuning of the parameters of the model must be introduced.

Another explanation could be provided by a modification of the laws of gravity on large scales. Such modifications generically involve a scalar degree of freedom which can lead to dynamical dark energy when the range of the scalar interaction is cosmological. As a result, scalar-tensor models with couplings to matter represent a well-motivated  and versatile class of dark energy. Theories describing the behaviour of the scalar field involve conformal \cite{Brans:1961sx} and disformal couplings to matter \cite{Bekenstein:1992pj,Zumalacarregui:2010wj}. It turns out that the disformal coupling is severely constrained by local experiments and cosmological observations \cite{Ip:2015qsa,Sakstein:2015jca}. On the other hand, the conformal couplings, albeit large on cosmological scales, can be screened in the local environment where none of their effects, such as deviations from Newton's law, have been uncovered. 

In this paper, we will focus on scalar-tensor models with screening mechanisms that are broadly classified to be of chameleon type \cite{Khoury:2003rn,Brax:2004qh}, {\it i.e.} where either the mass of the scalar and/or its coupling to matter has a dependence of the local matter density. Specifically, we will consider three types of models with the chameleon property: the $f(R)$ theories \cite{Hu:2007nk,Brax:2008hh}, the environmentally dependent Dilatons \cite{Brax:2010gi}  and the Symmetron \cite{Hinterbichler:2010es}. The latter two models use the Damour--Polyakov mechanism for screening \cite{Damour:1994zq}. We will  take advantage of the fact that these three very different types of models can be described using the same formalism defined in terms of two dynamical functions $m(a)$ and $\beta (a)$, where $a$ is the scale factor \cite{Brax:2011aw,Brax:2012gr}. The first one represents the mass of the scalar in the cosmological background at the redshift $1+z= a^{-1}$ and the second one is the coupling of the scalar to matter. The growth of cosmological perturbations in these models in the linear regime and on sub-horizon scales can be entirely described using a single function, $\epsilon (k,a)= 2 \beta^2(a)/[1+ m^2(a) a^2/k^2]$, which appears in the modification of Newton's constant and in the modified relation between the curvature and the gravitational potential. While in this paper we shall restrict ourselves to observables which are sensitive to the linear regime only, we note that, given $m(a)$ and $\beta(a)$, one can also reconstruct the full non-linear dynamics of the models. Namely, using a known evolution of the background matter density, $\rho(a)$, one can express the mass and the coupling as functions of local matter density: $m(\rho(x,t))$ and $\beta(\rho(x,t))$ and use them to perform N-body simulations of these models or to analyse local gravitational tests. 

Modified gravity (MG) and its comparison with dark energy has been investigated using various cosmological probes in the last ten years~\cite{Ishak:2005zs,Linder:2005in,Kunz:2006ca,Zhang:2007nk,Amendola:2007rr,Song:2008vm,Zhao:2008bn,Song:2008xd,Zhao:2009fn,Daniel:2010ky,Zhao:2010dz,Pogosian:2011op,Shapiro:2010si,Daniel:2010yt,Song:2010fg,Hojjati:2011ix,Hojjati:2011xd,Hojjati:2012ci,Hojjati:2012rf,Daniel:2012kn,Simpson:2012ra,Silvestri:2013ne,Hojjati:2013xqa,Hu:2013aqa,Dossett:2014oia,Ade:2015rim}. Some models of modified gravity, such as the $f(R)$ theories, have been strongly constrained by observations both cosmological and astrophysical. The strongest bound on the range of the scalar interaction, expressed in terms of the parameter $f_{R_0}$, is at the level of $10^{-7}$ and comes from astrophysical tests of modified gravity using the period of cepheids or the gas dynamics of dwarf galaxies~\cite{Jain:2012tn,Vikram:2013uba,Sakstein:2014nfa}. The cosmological bounds are less effective, at the level of $10^{-5}$~\cite{Dossett:2014oia,Bel:2014awa}. On the other hand, dilatons and symmetrons have not been constrained as systematically as $f(R)$ on cosmological scales. Only a few tests have been performed using the $[m(a),\beta(a)]$ parameterisation \cite{Hu:2013aqa}. The strongest bounds on dilatons and symmetrons still spring from local gravitational tests \cite{Brax:2013doa}. Local tests of gravity for the chameleon-type models of modified gravity imply that the range of the scalar interaction cannot exceed $1 \ {\rm Mpc}$, implying that linear analyses are limited to probing only some of the features of the chameleon screening mechanisms. On the other hand, studying effects of modified gravity on shorter scales requires the use of either semi-analytical methods suited to the quasi-linear regime of cosmological perturbations or N-body simulations, both of which are model-specific. Here, in order to keep our analysis as model-independent as possible, we shall restrict ourselves to observables which are sensitive to the linear regime only. 

The range of scales that are safely in the linear regime at low redshifts is quite limited. Most of the large scale structure data available today is from relatively low redshifts and provides only weak constraints on scalar-tensor models unless one considers information from non-linear scales. The only way to do so is to run N-body simulations for specific models. On the other hand, future surveys, such as LSST \cite{LSST} and Euclid \cite{Euclid}, will provide a high volume of data from higher redshifts at which the range of linear scales is significantly larger, allowing one to deduce stronger constraints on scalar interactions not only for specific models but in a more general model-independent way. In this paper, we start by deriving constraints on $f(R)$, Symmetron and Dilatons models from the subset of today's data that can be safely considered to be in the linear regime. Then we perform a Fisher forecast for the same models assuming data from a future LSST-like survey in combination with other types of data expected over the next 5-10 years to show that they will be significantly tighter. Finally, we perform a principal component analysis (PCA) forecast of $m(a)$ for the same future data, assuming that $\beta(a)$ is a slowly varying function that can be taken to be a ${\cal O}(1)$ constant over the range of redshifts relevant to LSST.

\section{The model}

We consider scalar-tensor theories defined by the action
\be
S=\int d^4 x \sqrt{-g}\left[ \frac{R}{16\pi G} + L_\phi+L_m [\psi, A^2(\phi) g_{\mu\nu}] \right] \ ,
\label{act}
\ee
where $g_{\mu\nu}$ is the Einstein frame metric, $\psi$ are the matter fields that follow geodesics of $A^2(\phi) g_{\mu\nu}$, and $L_\phi$ is the scalar field Lagrangian given by
\be
L_\phi = - \frac{(\partial \phi)^2}{2} -V(\phi) \ .
\label{lagrangian}
\ee
The action in Eq.~(\ref{act}) is a Generalized Brans-Dicke (GBD) theory \cite{Brans:1961sx} that includes a potential for the scalar field. In all GBD, the scalar field mediates an additional gravitational interaction between massive particles. The net force on a test mass is given by
\be
{\vec f}  = -{\vec \nabla} \Psi  -  \frac{d \ln A(\phi)}{d \phi} {\vec \nabla} \phi \ ,
\ee
where $\Psi$ is the Newtonian potential. Since solar system and laboratory tests severely constrain the presence of the scalar force, GBD can only be viable if either the coupling of the scalar field to matter is always negligible, or if there is a dynamical screening mechanism that suppresses the force in dense environments. The latter can be accomplished with appropriately chosen functional forms of $A(\phi)$ and $V(\phi)$. Because of its coupling to matter, the scalar field dynamics are determined by an effective potential which takes into account the presence of the conserved matter density $\rho$ of the environment
\be
V_{\rm eff}(\phi) =V(\phi) +(A(\phi)-1) \rho.
\ee
For some forms of $V(\phi)$ and $A(\phi)$, the effective potential can have a density dependent minimum,
$\phi_{}(\rho)$.  The scalar force will be screened if either the mass of the field happens to be extremely large or the coupling happens to be negligibly small at the minimum of $V_{\rm eff}(\phi)$. Such models can be broadly classified as ``Generalized Chameleons'' (GC), and include the original chameleon model \cite{Khoury:2003rn}, $f(R)$,  dilatons  \cite{Brax:2010gi} and  symmetrons \cite{Hinterbichler:2010es}.

We note that the GC scalar-tensor theories considered in this work are viable only if the field stays at the minimum of the effective potential $V_{\rm eff}(\phi)$ \cite{Brax:2012gr}. In this case, the effective dark energy equation of state is indistinguishable from $-1$ and the expansion history practically the same as in the $\Lambda$CDM model. Furthermore, as long as the scalar field is at its density dependent minimum, $\phi (\rho)$, the theory can be described parametrically from the sole knowledge of the mass function $m(\rho)$ and the coupling $\beta (\rho)$ at the minimum of the potential \cite{Brax:2012gr,Brax:2011aw}
\be
\frac{\phi (\rho)-\phi_c}{m_{\rm Pl}}= \frac{1}{m_{\rm Pl}^2}\int_{\rho}^{\rho_c} d\rho \frac{\beta (\rho) }{m^2(\rho)},
\ee
where we have identified the mass as the second derivative
\be
m^2 (\rho)= \frac{d^2 V_{\rm eff}}{d\phi^2}\vert_{\phi=\phi (\rho)}
\ee
and the coupling
\be
\beta (\rho)= m_{\rm Pl} \frac{d\ln A}{d\phi}\vert_{\phi=\phi(\rho)}.
\ee
It is often simpler to characterize the functions $m(\rho)$ and $\beta(\rho)$ using the time evolution of the matter density of the Universe
\be
\rho(a)=\frac{\rho_0}{a^3}
\ee
where $a$ is the scale factor whose value now is $a_0=1$. This allows one to describe characteristic models in a simple way and the full dynamics can be recovered from the time evolution of the mass and coupling functions, $m(a), \beta(a)$.

\subsection{Evolution of linear perturbations}

While the scalar-tensor theories considered in this work predict the same expansion history as $\Lambda$CDM, the existence of the additional scalar interaction gives them distinguishing features in the evolution of linear matter and metric perturbations. More specifically, the attractive force mediated by the scalar enhances the overall growth of inhomogeneities. In addition, the relation between the curvature perturbation $\Phi$ and the Newtonian potential $\Psi$ is modified \cite{Schimd:2004nq}. Both of these effects can be captured in terms of two phenomenological functions employed in MGCAMB\footnote{MGCAMB is a publicly available patch to CAMB \cite{CAMB}.} \cite{Zhao:2008bn,Hojjati:2011ix,MGCAMB}, parametrizing effective modifications to the Poisson and the anisotropy Einstein equations in Fourier space. Namely, one defines $\mu(a,k)$ and  $\gamma(a,k)$, such that 
\begin{gather}
k^2\Psi = -4\pi Ga^2\mu(k,a)\rho\Delta, \\
 \frac{\Phi}{\Psi} = \gamma(k,a)
\end{gather}
where $\Delta$ is the comoving matter density contrast\footnote{These equations are valid at late times, when the contribution of relativistic species can be neglected. Equations used in MGCAMB are more general and are valid at all times \cite{Hojjati:2011xd}.}. In the quasi-static approximation, whose validity is discussed below, functions $\mu(k,a)$ and $\gamma(k,a)$ can be expressed in terms of $m(a)$ and $\beta(a)$ as \cite{Brax:2012gr}.
\begin{gather}
\label{mofmb}
\mu(a,k) = A^2(\phi)(1+\epsilon(k,a)) , \\
\label{gofmb}
\gamma(a,k) = \frac{1-\epsilon(k,a)}{1+\epsilon(k,a)} \ ,
\end{gather}
where 
\be
\epsilon(k,a)= \frac{2\beta^2(a)}{1+ m^2(a) a^2/k^2} \ .
\label{epsilon}
\ee
The conformal factor $A^2(\phi)$ that appears in Eq.~(\ref{mofmb}) is indistinguishable from unity for viable models within the class of scalar-tensor theories considered in this paper, and can be safely ignored. $\Lambda$CDM is recovered when $\epsilon \rightarrow 0$ and $\mu=\gamma=1$. 

In the quasi-static approximation, the equation governing the evolution of matter density contrast $\delta$ reads
\be
\delta'' + {\cal H} \delta' -\frac{3}{2} \Omega_m {\cal H}^2 \mu (k,a) \delta=0
\label{gt}
\ee
where $'$ is the derivative with respect to conformal time and ${\cal H}=a'/a$. Two regimes can be distinguished. When the mode $k$ is outside the Compton wavelength of the scalar field, i.e. $k\ll a m(a)$, $\epsilon \ll 1$ and the growth is not modified. Inside the Compton wavelength, $k\gg a m(a)$, gravity is enhanced by $1+2\beta^2(a)$, implying more growth. In addition, in the Symmetron and Dilatons models, the coupling $\beta(a)$ depends on the matter density and controls the transition to the enhanced growth. 

Since $\epsilon$ is a manifestly non-negative number, the growth is generically enhanced. Also, generically, $\gamma < 1$ in these models. At the same time, the relation between the lensing potential $\Phi+\Psi$ and the matter density is effectively unchanged. Namely, if one defines $\Sigma(k,a)$ as
\be
k^2(\Psi + \Psi) = -8\pi Ga^2 \Sigma(k,a) \rho\Delta, \\
\ee
then 
\be
\Sigma = A^2(\phi)
\ee
and is effectively unity for all viable modes. Thus, a clear detection of $\Sigma \ne 1$ would not only signal a breakdown of $\Lambda$CDM but would rule out the entire class of GBD models. We note that, even though $\Sigma$ is constrained to be very close to unity in viable GBD models, its time derivative, $\dot{\Sigma}$, can, in principle, be non-negligible and affect the observables via the Integrated Sachs-Wolfe (ISW) effect.

When functions $m(a)$ and $\beta(a)$ are regular, which is the case for chameleon models such as $f(R)$, and for dilatons, the error introduced by working in the quasi-static approximation scales as $H/k$ \cite{Brax:2013fna}. For models such as the Symmetron, in which the functions $m(a)$ and $\beta(a)$ vanish with a power $n<1$ for $a>a_\star$ and are zero for $a<a_\star$ (and thus have a diverging derivative at $a_\star$), the accuracy is reduced to $(H/k)^n$ \cite{Brax:2013fna,Llinares:2013qbh}.

\subsection{Functions $m(a)$ and $\beta(a)$ in f(R)}

\begin{figure*}[tbp]
\begin{center}
\includegraphics[width=0.95\textwidth]{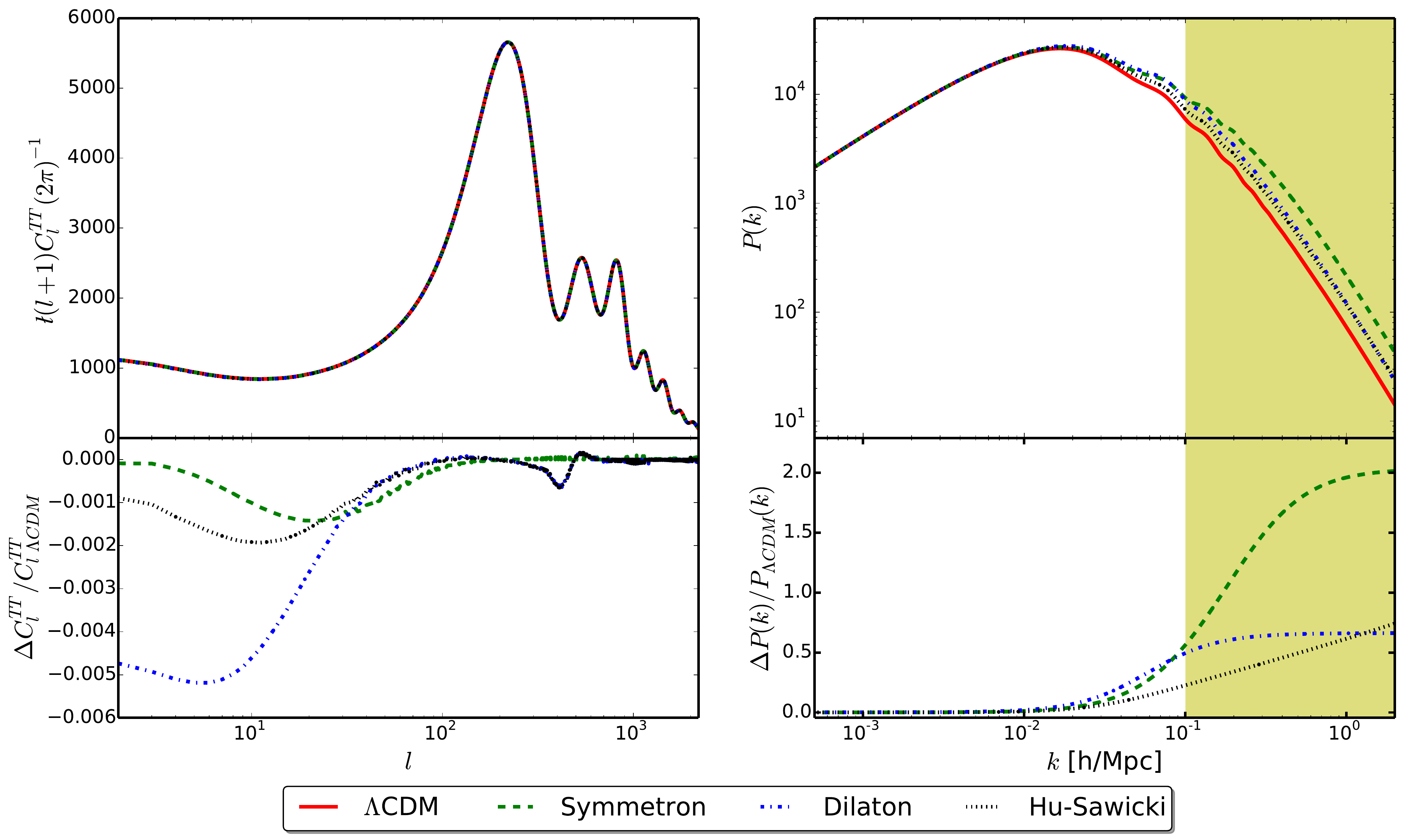}
\caption{
Plots of CMB temperature anisotropy $C^{TT}_l$ (left) and the matter power spectrum $P(k)$ (right) for the models studied in this paper. The parameters used for the Symmetron model are: $a_{\star}=0.25$, $\beta_{\star} = 1$ and $\xi_{\star}=10^{-3}$. The parameters used for the Dilatons model are: $\beta_0 = 3$ and $\xi_0=6 \times 10^{-3}$. The parameters used for Hu-Sawicki $f(R)$ model are $f_{R_0}=10^{-4}$ and $n=1$. The yellow shaded region shows the scales that are not taken into account in the data analysis.}
\label{fig:samples}
\end{center}
\end{figure*}

In what follows, we briefly motivate specific functional forms of $(m(a)$ and $\beta(a)$ adopted for the analysis in Sections \ref{sec:constraints} and \ref{sec:forecasts}. Given the forms of $m(a)$ and $\beta(a)$, the predictions for the observables can be calculated using MGCAMB \cite{MGCAMB}. Plots of the CMB temperature anisotropy and the matter power spectra for a few representative models are shown in Fig.~\ref{fig:samples}. 

Among theories exhibiting chameleon screening are the $f(R)$ class of models \cite{Capozziello:2003tk,Carroll:2003wy} described by the action
\be
S=\int d^4 x \sqrt{-g}\left[ \frac{f(R)}{16\pi G} +L_m [\psi, g_{\mu\nu}] \right]
\ee
where the function $f(R)$ is designed to depart from the Einstein-Hibert form at smaller values of the curvature $R$. As a specific example, we take the form proposed by Hu and Sawicki (HS) \cite{Hu:2007nk},
\be
f(R)=R - 2\Lambda + \frac{f_{R_0}}{n} \frac{R_0^{n+1}}{R^n} \ ,
\label{frn}
\ee
where $\Lambda$ is the cosmological constant term, $R_0$ is the value of the curvature today and $f_{R_0} \equiv (1-df/dR)_{R=R_0}$. As argued in \cite{Hu:2007nk,Appleby:2007vb,Starobinsky:2007hu}, all viable $f(R)$ models should be of such ``disappearing cosmological constant'' type \cite{Starobinsky:2007hu}, and models similar to HS were proposed in \cite{Appleby:2007vb,Starobinsky:2007hu}. 

For all $f(R)$ models, $\beta(a)=1/\sqrt{6}$, while the mass function is model dependent. In the HS model, we have
\be
\label{eq:fR_m}
m(a)= m_0 \left(\frac{4\Omega_{\Lambda}+ \Omega_{m} a^{-3}}{4\Omega_{\Lambda}+ \Omega_{m}}\right)^{(n+2)/2}
\ee
where $\Omega_{\Lambda}$ and $\Omega_{m}$ are the dark energy and matter density fractions today, and $m_0$ is a mass scale that can be expressed in terms of $f_{R_0}$ as \cite{Brax:2012gr}
\be
\label{eq:fR_m0}
m_0= H_0 \sqrt{\frac{4\Omega_{\Lambda}+ \Omega_{m} }{(n+1) f_{R_0}}}.
\ee
Local tests of gravity require $f_{R_0}\lesssim 10^{-6}$  \cite{Brax:2013doa}, while astrophysical constraints from dwarf galaxies imply that $f_{R_0}\lesssim  10^{-7}$ \cite{Sakstein:2014nfa}. These bounds depend on accurate modelling of non-linear physics. In what follows, we will derive the constraint on $f_{R_0}$ from current cosmological data using only information from linear scales, and also forecast constraints expected from future surveys like LSST. 

Representative CMB and matter power spectra for $f(R)$ are shown in Fig.~\ref{fig:samples}. A notable effect on the CMB spectrum is the suppression of power at small multipoles, which is due to the reduced Integrated Sachs-Wolfe (ISW) effect. The magnitude of the ISW effect is proportional to the net change in the gravitational potential along the line of sight. In $\Lambda$CDM, the change in the potential is a reduction caused by the onset of cosmic acceleration. In $f(R)$, the additional scalar force enhances the potential which, combined with the decay due to acceleration, leads to a smaller net change and, thus, a smaller ISW effect. The other notable impact of $f(R)$ on the CMB spectrum is the enhanced lensing, which has the effect of slightly dumping the peaks. The enhanced growth is more evident in the plot of $P(k)$. Qualitatively, these features are common to all GBD models.

\subsection{Functions $m(a)$ and $\beta(a)$ for dilatons}

Another relevant example is the environmentally dependent Dilaton \cite{Brax:2010gi}, where the screening mechanism is of the Damour-Polyakov type \cite{Damour:1994zq}. This model, inspired by string theory in the large string coupling limit,
has an exponentially runaway potential
\be
V(\phi)=V_0 e^{-\phi/m_{\rm Pl}} \ ,
\ee
with the value of $V_0$ set to generate the current acceleration of the Universe, while the coupling function is
\be
A(\phi)=1+\frac{A_2}{2m_{\rm Pl}^2} (\phi-\phi_\star)^2 \ .
\ee
In dense environments, the minimum of the effective potential approaches $\phi = \phi_\star$, and the coupling function $\beta(a)$ vanishes. The coefficient $A_2$ has to be large to satisfy local tests of gravity; typically $A_2 \gtrsim 10^6$.
These models can be described by a mass function given by
\be
\label{eq:dilaton_m}
m^2(a)= 3 A_2 H^2(a)
\ee
and, assuming matter domination, a coupling function
\be
\beta(a)= \beta_0 a^3 \ ,
\ee
where $\beta_0=\Omega_{\Lambda}/ \Omega_{m} \sim 2.7$ is related to $V_0$, and is determined by requiring that $\phi$ plays the role of dark energy.  We will present our constraints on the mass in terms of a scalar-force range parameter $\xi_0$, defined as 
\begin{equation}
\xi_0 = \frac{H_0}{c~m_0} = \frac{1}{ \sqrt{3 A_2}} ,
\label{eq:xi0}
\end{equation}
where $m_0=m(a=1)$. 
We show representative CMB and matter power spectra for the Dilaton model in Fig.~\ref{fig:samples}, with parameter values being large on purpose to exaggerate the qualitative features of the model.

\subsection{Functions $m(a)$ and $\beta(a)$ for symmetrons}

Another example of a GBD model with the Damour-Polyakov screening mechanims is the Symmetron \cite{Hinterbichler:2010es}, where the scalar field has a quartic potential,
\be
V(\phi)= V_0 + \frac{m_\star^2\phi_\star^2}{2}\left[ - {1\over 2} \left({\phi \over 2\phi_\star}\right)^2 + {1\over 4} \left({\phi \over \phi_\star}\right)^4 \right]
\ee
and a coupling function,
\be
A(\phi)= 1 + \frac{\beta_\star}{2\phi_\star} \phi^2 \ .
\ee
When matter density is large, the effective potential has a minimum at $\phi=0$ and $A(\phi) \rightarrow 1$, thus decoupling the scalar from matter. At lower densities, the effective potential acquires a non-zero minimum, activating the scalar force. For cosmological densities, the transition occurs at 
\be
\rho_\star= {\rho_{m}\over a_\star^3}={m_{pl}m_\star^2\phi_\star^2\over 2\beta_\star} \ ,
\ee 
where $\rho_m$ is the matter density today. Thus, one can work with $a_\star$, along with $m_\star$ and $\beta_\star$, as the three free parameters of the theory. At $a>a_\star$, the model can be described by 
\be
m(a)=m_\star \sqrt{1- \left(\frac{a_\star}{a} \right)^3}
\ee
and
\be
\beta(a)=\beta_\star \sqrt{1-\left(\frac{a_\star}{a}\right)^3} \ ,
\ee
while $\beta (a)=0$ for $a<a_\star$. As in the case of dilatons, we represent our bounds in terms of a range parameter $\xi_\star$, defined as
\begin{equation}
\xi_\star = \frac{H_0}{c} \frac{1}{m_{\star}} .
\label{eq:xistar}
\end{equation}
Representative CMB and matter power spectra for this model are shown in Fig.~\ref{fig:samples}.

\subsection{Generalized Chameleon models}

In our forecasts, we will also consider generalized models of chameleon type \cite{Brax:2013mua} defined by
\be
m(a)= m_0 a^{-r},\ \beta(a)= \beta_0 a^{-s} \ .
\ee
In practically all viable chameleon models, the coupling function is expected to vary extremely slowly at redshifts probed by large scale structure surveys. Thus, for all practical purposes, it can be taken to be a constant of order unity. 

\subsection{Binned Model}
\label{sec:binnedmodel}

As discussed so far, for any of the aforementioned models, each with its own theoretical motivation, one can determine the functional forms of $m(a)$ and $\beta(a)$. This effectively reduces the two free functions $m(a)$ and $\beta(a)$ to a handful of parameters. However, one might be interested in knowing how well the two functions are constrained in general, without regard for any specific model. One can then proceed by discretizing either of the two functions in bins of redshift space and treating the amplitude in each bin as a free parameter to be constrained. 

Varying both, the coupling and the mass functions, simultaneously would be redundant, since their effect is largely degenerate. Since it is the mass parameter that affects the shape of the matter power spectrum, we fix $\beta(a)$ to a constant value of order unity and bin $m(a)$ in redshift. If a non-zero $m^{-1}(a)$ were detected, it would signal the presence of a scalar interaction and further investigation would be required to determine if the variation occurs in $\beta(a)$, $m(a)$ or both. 

While a binning scheme gives a model independent (rather a far less model dependent) treatment of $m(a)$, the larger number of parameters (values of $m$ in each bin) results in weaker constraints on the individual parameters. To extract useful information, we apply the Principal Components Analysis (PCA) technique (reviewed in Section \ref{PCA}). The resulting Principal Components (PCs) are linear combinations of the original bin values and the propagated uncertainty (from original errors on the bins) in their values can inform us about those PCs that are best constrained by data and the number of degrees of freedom the can potentially be constrained.  

\section{Constraints from current data}
\label{sec:constraints}

In this Section, we use a combination of currently available CMB, lensing and Baryonic Acoustic Oscillation (BAO) data, as well as measurements of the matter power spectrum, to derive constraints on the GBD parameters.  To compute the observables, we implemented the parametrizations described in the previous Section in MGCAMB. We then use it with an appropriately modified version of CosmoMC \cite{cosmomc} to obtain the posterior distributions for the model parameters. Since current data is unable to simultaneously constrain multiple GBD parameters, we will only consider models from the previous Section for which meaningful constraints are possible.

\subsection{The datasets used in the analysis}
\label{sec:current_data}
We use the measurements of CMB temperature anisotropy from the second data release of the \emph{Planck} survey  \cite{PlanckResult2015:CosmoParams} in the form of the full \emph{Planck} TT high-$\ell$ likelihood ($30<\ell<2500$) along with the low-$\ell$ polarization ($\ell <30$). We refer to the above datasets as PLC. We also consider the \emph{Planck} 2015 lensing potential spectrum \cite{PlanckLensing} extracted from mode-coupling correlations, and refer to this dataset as CMBLens. 

In addition to inducing higher order correlations, lensing by large scale structures affects the TT spectrum at higher $\ell$, slightly damping the oscillatory features. In \cite{Ade:2013zuv}, and subsequently in \cite{PlanckResult2015:CosmoParams}, the lensing contribution to TT was quantified via an amplitude $A_L$ multiplying the lensing power spectrum in the calculation of the theoretical prediction for TT. The parameter $A_L$ was used to quantify the significance of detection of the lensing contribution to TT. However, instead of measuring the expected value of  $A_L=1$, since the lensing contribution to TT is calculated from the same model as the rest of the spectrum, the best fit value obtained for LCDM from the PLC dataset in \cite{PlanckResult2015:CosmoParams} was $A_L=1.22 \pm 0.10$, or two standard deviations away from the expectation. As discussed in \cite{Ade:2013zuv} (see also \cite{Addison:2015wyg}) this is due to an apparent tension between the higher-$\ell$ and lower-$\ell$ data when trying to fit LCDM to Planck TT data. To negate the effect of this tension, the parameter $A_L$ was sometimes co-varied with other parameters when deriving constraints on LCDM in \cite{PlanckResult2015:CosmoParams}. In what follows, we take the view that $A_L$ is not a physical parameter and should be held fixed to 1 when deriving constraints on cosmological models. However, we also investigate and discuss the effect of co-varying $A_L$ in the case of $f(R)$.

For BAO measurements, we used data from the 6dF survey \cite{6dFBAO} and from SDSS, specifically the MGS \cite{SdssBaoMGSS} and BOSS data releases (LOWZ  and CMASS) \cite{BossBAO}. 

We also use the matter power spectrum (referred to as MPK) from SDSS LRG DR4 \cite{Tegmark06}, but only on linear scales, $k \le 0.1 \, h/\text{Mpc}$. We are aware of the fact that non-linear corrections can play a role even at $k \lesssim 0.1$Mpc and that a proper treatment of the bias and the redshift space distortions (RSD) must take them into account. This was studied at length in \cite{Zhao:2012xw} for the SDSS DR9 power spectrum and it was found that the differences in the upper bounds on neutrino masses obtained using four different RSD models were under $20$\%. Based on this, we expect that bounds on the GBD parameters (such as $f_{R_0}$) obtained from MPK are accurate to within $30$\%, which is sufficient given that constraints form current data are relatively weak.

Finally, we consider the weak lensing data from the Canada France Hawaii Telescope Lensing Survey 2DCFHTLenS \cite{Kilbinger2013}, referred to as WL. To avoid dealing with non-linear scales, we adopt a conservative cut and exclude $\theta < 30^{\prime}$ from the measurements of the correlation function $\xi^{\pm}$, which corresponds to $k < 0.1 \, h/\text{Mpc}$ scales.

\subsection{Constraints on $f(R)$}

\begin{figure}[tbp]
\includegraphics[width=0.5\textwidth]{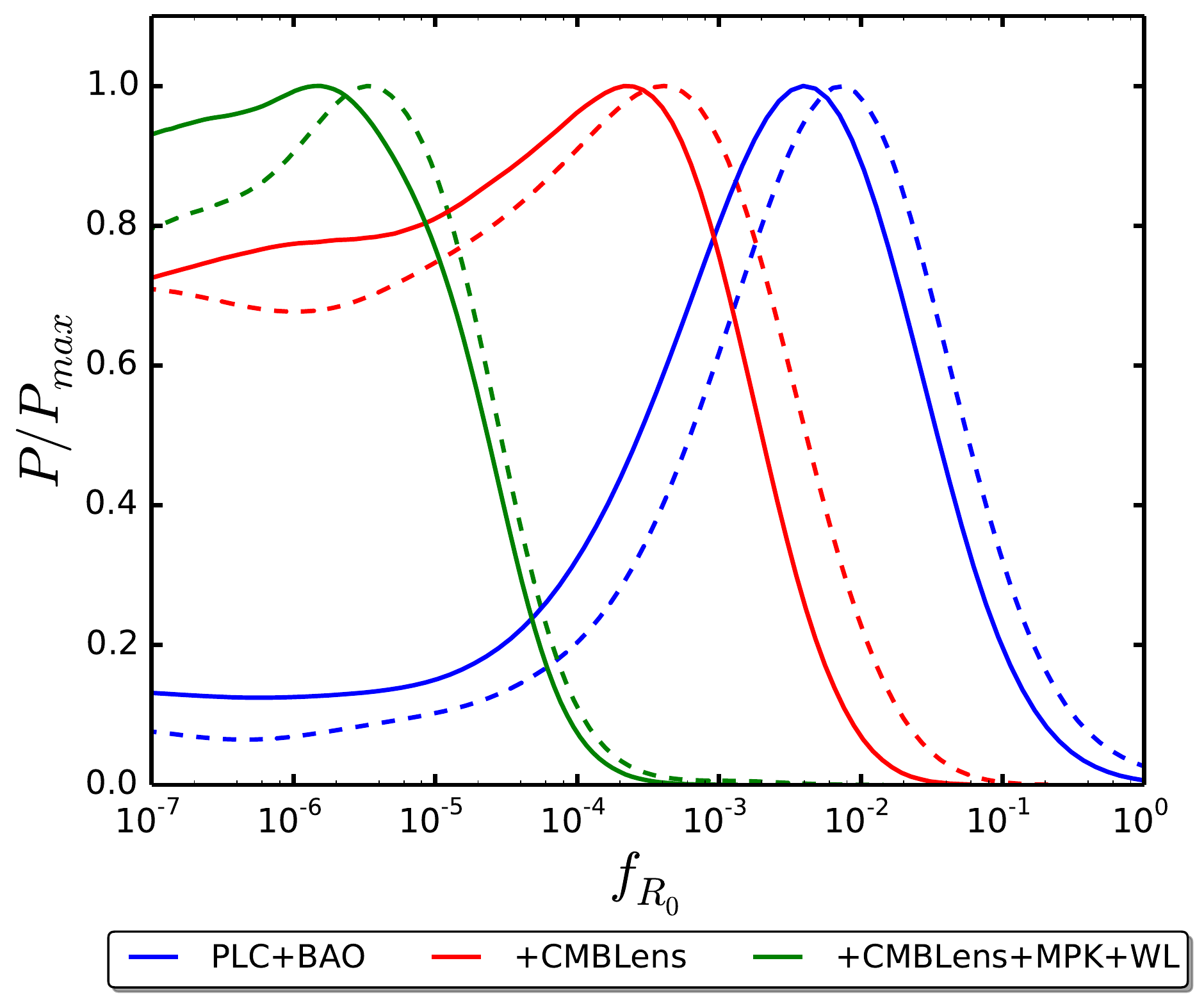}
\caption{\label{fig:husaw_dataset}
The marginalized posterior distribution for the $f_{R_0}$ parameter in the Hu-Sawicki model ($n=1$) for different combinations of datasets. The solid lines show the PDF in case of massive neutrinos with a fixed mass $\sum m_{\nu}=0.06 \, \text{eV}$, while the dashed lines show the PDF for the case when the neutrino mass was varying. Due to the degeneracy between $f_{R_0}$ and $\sum m_{\nu}$, we see that the constraint on $f_{R_0}$ become weaker when the neutrino mass is varied. The datasets are labeled according to the notation introduced in Sec.~\ref{sec:current_data}.  The symbol +  means that we add data on top of the PLC+BAO dataset. For example, +lensing means PLC+BAO+lensing.}
\end{figure}

\begin{figure}[tbp]
\includegraphics[width=0.47\textwidth]{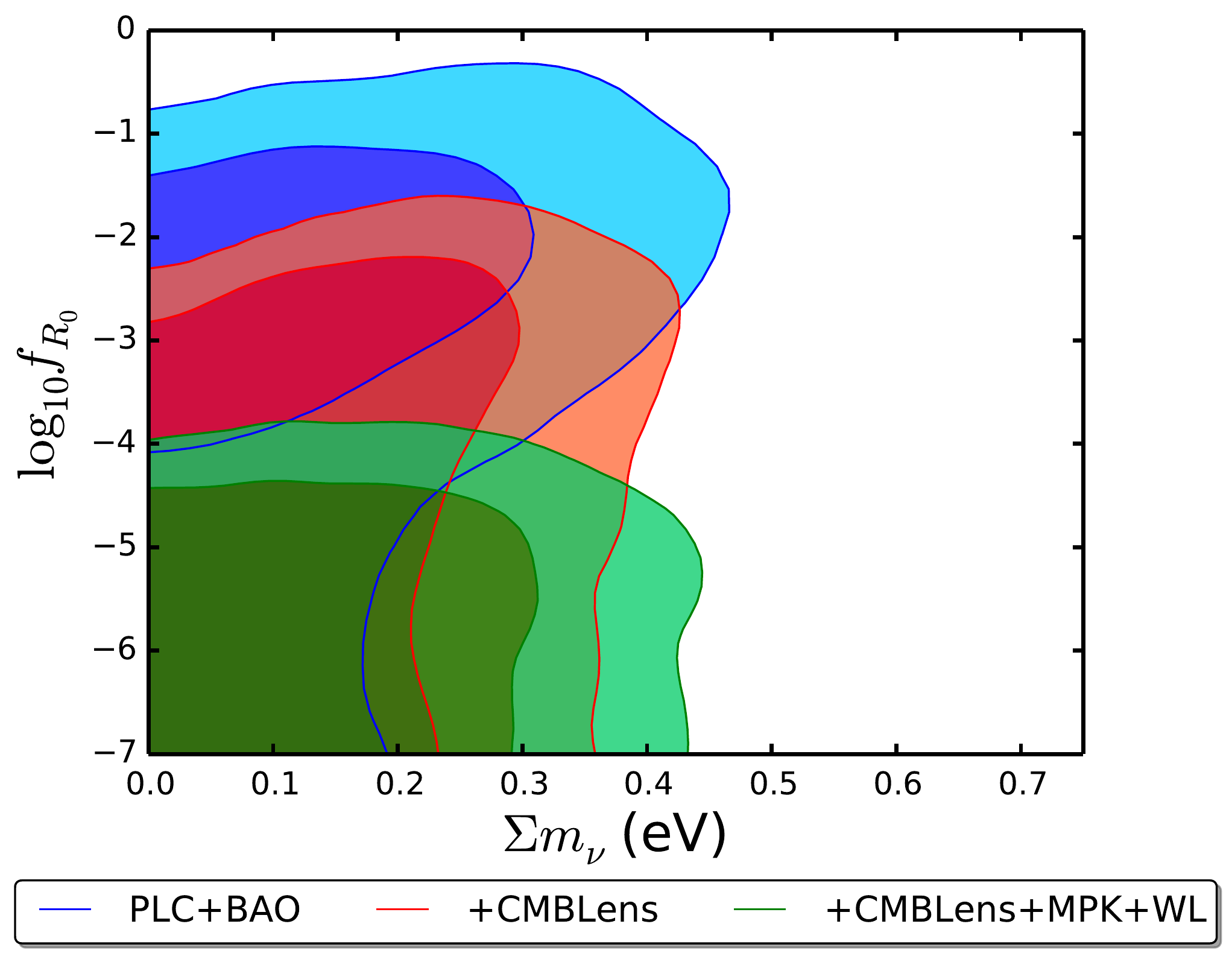}
\caption{\label{fig:husawnu}
Joint contours for $f_{R_0}$ and $\sum m_{\nu}$ in the Hu-Sawicki model ($n=1$) after marginalizing over all other cosmological parameters. The darker and lighter shades correspond respectively to the 68\% C.L. and the 95\% C.L. . Datasets are described in text and also in the caption of Fig.~\ref{fig:husaw_dataset}}
\end{figure}

\begin{figure}[tbp]
\begin{center}
\includegraphics[width=0.45\textwidth]{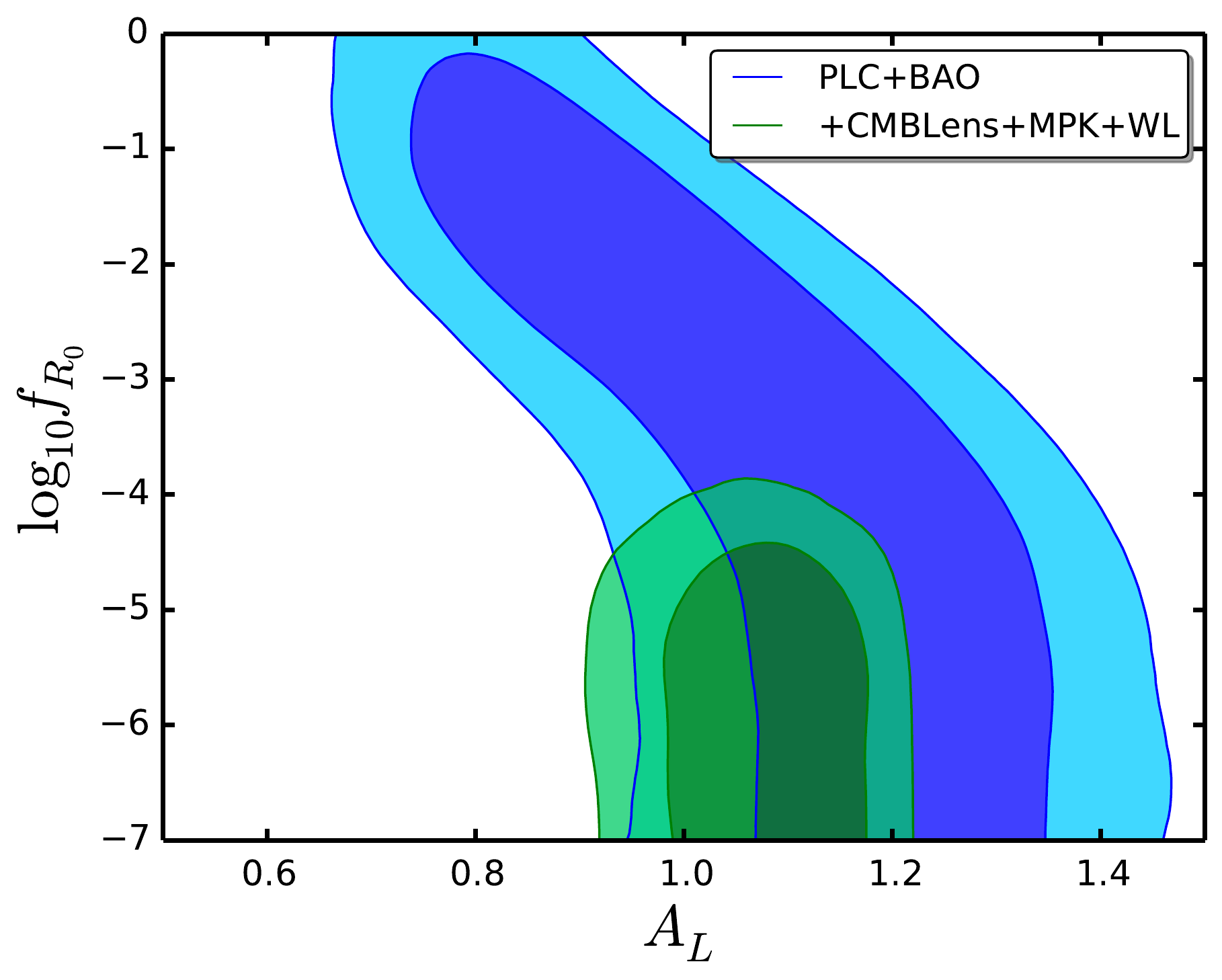}
\caption{\label{fig:husawAlens}
Joint contours for $f_{R_0}$ and $A_L$ in the Hu-Sawicki model. The darker and lighter shades corresponf respectively to the 68\% C.L. and the 95\% C.L. Using PLC+BAO data sets only it is possible to detect high values of $f_{R_0}$ that can cure the tension in lensing amplitude $A_L$. However such high values are ruled out once we add lensing and LSS data sets.}
\end{center}
\end{figure}

\begin{table}[tbp]
\centering
\begin{tabular}{l||c||c|c}
\hline 
  & Fixed $\sum  m_{\nu}$ & \multicolumn{2}{|c}{Varying $\sum m_{\nu}$ }  \\
\hline
 Data sets & $f_{R_0}$  & $f_{R_0}$  & $\sum m_{\nu}$ (eV) \\
  \hline
  PLC+BAO & $0.05 \,(0.14)$ & $0.08 \,(0.23)$ &  $0.24 (0.35)$ \\ 
  +CMBLens & $3 \,(8) \times 10^{-3}$ & $0.6 \,(1.6) \times 10^{-2}$ & $0.22 (0.31)$ \\
  +MPK & $0.6 (1.6) \times 10^{-4}$ & $0.7 \,(1.7) \times 10^{-4}$ & $0.24 (0.34)$ \\
  +WL & $3 \,(7) \times 10^{-5}$ & $4 \, (9) \times 10^{-5}$  & $0.23 (0.33)$\\
  \hline
  \end{tabular}
  \caption{The 68\% (95\%) CL upper limits of $f_{R_0}$ and the sum of neutrino masses using different combinations of data sets shown in the table.}
  \label{tab:husawdata}
\end{table}

\begin{table*}[tbp]
\centering
\begin{tabular}{c|c||c|c|c}
\hline
\multicolumn{2}{c||}{$f(R)$+$A_L$, fixed $\sum m_{\nu}$ } & \multicolumn{3}{|c}{$f(R)$+$A_L$, varying $\sum m_{\nu}$} \\
\hline
$f_{R_0}$  &  $A_L$ & $f_{R_0}$  &  $A_L$ & $\sum m_{\nu}$ \\
\hline
$3 \,(8)  \times 10^{-5}$ & $1.08^{+ 0.07\, (0.12)}_{-0.05 \, (0.13)}$ & $0.4 \, (1.0) \times 10^{-4} $& $1.11^{+0.10 \,(0.16)}_{-0.06 \, (0.15)}$ &  $0.30 \, (0.38)$\\
\hline
\end{tabular}
\caption{\label{tab:husaw_AL} 68\% (95\%) CL bounds on $f_{R_0}$, $A_L$ and $\sum m_{\nu}$ using all the data sets: PLC+BAO+CMBlens+MPK+WL }
\end{table*}

The Hu-Sawicki $f(R)$ model has two parameters, $f_{R_0}$ and $n$. In what follows, we fix $n=1$ because that is a common choice in the literature, and also because the two parameters are highly correlated and the current data cannot simultaneously constrain both. We chose a flat prior on $\log_{10} f_{R_0}$ within the $[-7,0]$ range. We have checked that changing the range of the flat prior does not affect our results.

Fig.~\ref{fig:husaw_dataset} shows constraints on $f_{R_0}$ for different combinations of datasets described in Sec.~\ref{sec:current_data}, after marginalizing over all the other cosmological parameters. We considered the case in which the total neutrino mass is fixed at $\sum m_{\nu}=0.06 \, \text{eV}$ (solid lines), and the case where it can vary within $0\le \sum m_{\nu} \le 1 \,\text{eV} $ (dashed lines). The results from Fig.~\ref{fig:husaw_dataset} are summarized in Table~\ref{tab:husawdata}. 

We can see that the combination of PLC and BAO datasets (blue lines) only weakly constrains the model. Modified gravity affects the CMB temperature anisotropy spectrum in two ways: it affects the low-$\ell$ power spectrum through the ISW effect  and enhances the damping at high-$\ell$ due to the enhancement in clustering and, as a consequence, the lensing potential. Thus, the observed lack of power at low-$\ell$ multipoles and the apparent preference of enhanced lensing in CMB TT, when compared to the $\Lambda$CDM prediction, can be reconciled by a non-zero $f_{R_0}$. This is the reason for the peak in the PLC+BAO likelihood. Adding the CMBLens data (red lines) tightens the constraint substantially. The enhancement of growth due to the extra scalar interaction affects the lensing potential measured by \emph{Planck}, which is known to be in excellent agreement with the LCDM prediction \cite{PlanckLensing}. Thus, the weak preference for larger $f_{R_0}$ coming from PLC+BAO is overwhelmed by the stronger CMBLens data that is consistent with $f_{R_0}=0$. The constraint becomes even tighter after adding the MPK and WL datasets (green lines).

The dashed lines in Fig.~\ref{fig:husaw_dataset} show the impact of co-varying the combined mass of neutrinos, $\sum m_{\nu}$, along with $f_{R_0}$. Massive neutrinos suppress the growth and can partially compensate for the enhanced clustering in $f(R)$,  slightly weakening the bounds on $f_{R_0}$. The extent of the degeneracy can be inferred from Fig.~\ref{fig:husawnu} which shows the joint confidence contours for the two parameters. We see that, although the constraint on $f_{R_0}$ becomes tighter as we add the LSS data, the constraint on $\sum m_{\nu}$ remains roughly the same. This is because we are restricting our analysis to linear scales, while the effect of massive neutrinos becomes more relevant on smaller scales and, hence, causes only a small degradation of $f_{R_0}$ constraints.

Up to this point, we kept the unphysical lensing amplitude parameter $A_L$ fixed at its expected value of 1. However, one may wonder if the discrepancy in $A_L$ observed in the $\Lambda$CDM model also persists in $f(R)$, and what effect co-varying $A_L$ has on the bounds on $f_{R_0}$. The results for two different combinations of data are shown in Fig.~\ref{fig:husawAlens}. 
Although it seems that, in the case of PLC+BAO, the lensing amplitude tension has been reconciled, we argue that this is not due to a genuine signal of modified gravity. As discussed previously, the PLC+BAO data yields a peak in the likelihood of $f_{R_0}$ because the preference for enhanced lensing and the lack of power at low $\ell$ in $C_{\ell}^{TT}$ can be reconciled with a non-zero $f_{R_0}$. The enhanced lensing appears to cure the $A_L$ problem and this is depicted in Fig.~\ref{fig:husawAlens}, where we see that there is a strong degeneracy between $A_L$ and $f_{R_0}$ for large values of the latter (blue contours). However such large values of $f_{R_0}$ are ruled out once we add the datasets that probe clustrering (green contours). Still, the value of $A_L$ when co-fit with $f_{R_0}$ is in better disagreement with the prediction. For the combination of all data we find 
\be
A_L= 1.11^{+0.20}_{-0.14} \quad \text{68 \% C.L., all datasets}.
\ee
The results of the analysis with varying $A_L$ are summarized in Table~\ref{tab:husaw_AL}.

\subsection{Constraints on the Symmetron model}

\begin{figure}[tbp]
\includegraphics[width=0.5\textwidth]{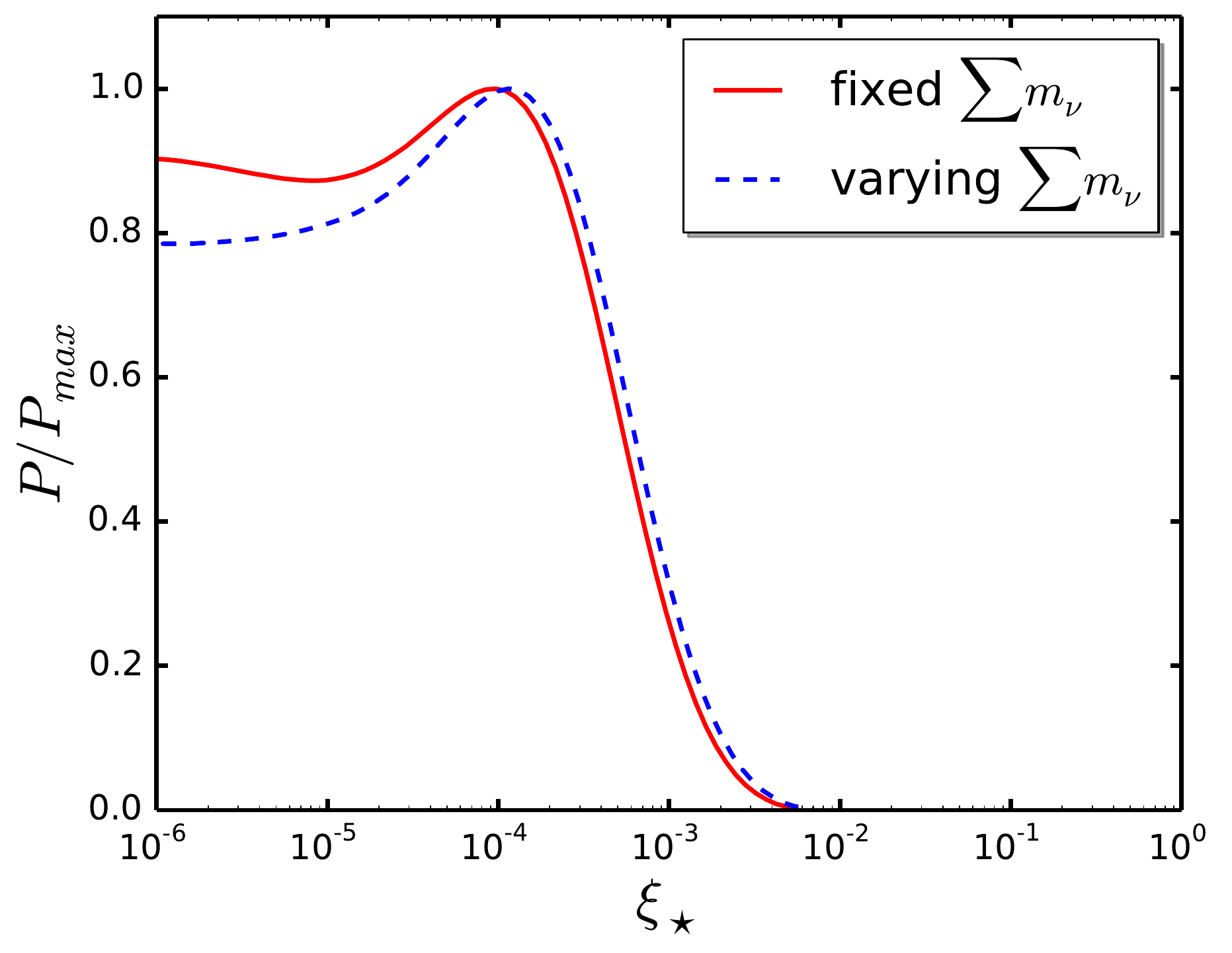}
\caption{\label{fig:symmetron1D}
The marginalized posterior distribution for $\xi_{\star}$ in Symmetron model with $\beta_{\star}=1$ and $a_{\star} = 0.25$ considering neutrinos with $\sum m_{\nu} = 0.06$ eV (red solid line) and marginalizing over a varying $\sum m_{\nu}$ (blue dashed line). The data sets used in this analysis are PLC+BAO+lensing+MPK+WL as described in section \ref{sec:current_data}.}
\end{figure}

In this Subsection we derive constraints on the inverse mass parameter, $\xi_\star$, defined in Eq.~(\ref{eq:xistar}), which represents the Compton wavelength of the scalar interaction. We fix the other two Symmetron parameters, taking $a_{\star} = 0.25$ and $\beta_{\star} = 1$, since current data is unable to constrain them simultaneously with $\xi_\star$.

Fig.~\ref{fig:symmetron1D} shows the posterior probability distribution for the $\xi_{\star}$ parameter with a fixed $\sum m_{\nu} = 0.06$ eV (red solid line) as well as after marginalizing over a varying $\sum m_{\nu}$ (blue dashed line).  We find an upper bound of $\xi_{\star} < 1.5 \times 10^{-3}$ at 95 \% C.L, which corresponds to a Compton wavelength of $\sim$ a few Mpc.  Our bounds are summarized in Table~\ref{tab:summ_fit}.

As mentioned above, current data is unable to simultaneously constrain all the model parameters because they are highly correlated. We also note that one cannot derive meaningful constraints for smaller values of coupling constant $\beta_{\star}$ as the modification of growth is relatively small for the scales and redshifts currently probed. Further, since $a_{\star}$ sets the onset of modified growth,  we would see tighter constraints on $\xi_{\star}$ for smaller $a_{\star}$ values. Nevertheless, as we will show in Sec.~\ref{sec:forecasts}, future surveys with larger sky and deeper redshift coverage will be able to constrain $\xi_{\star}$ along with the other two parameters. 

\subsection{Constraints on the Dilaton model}

\begin{figure}[tbp]
\begin{center}
\includegraphics[width=0.5\textwidth]{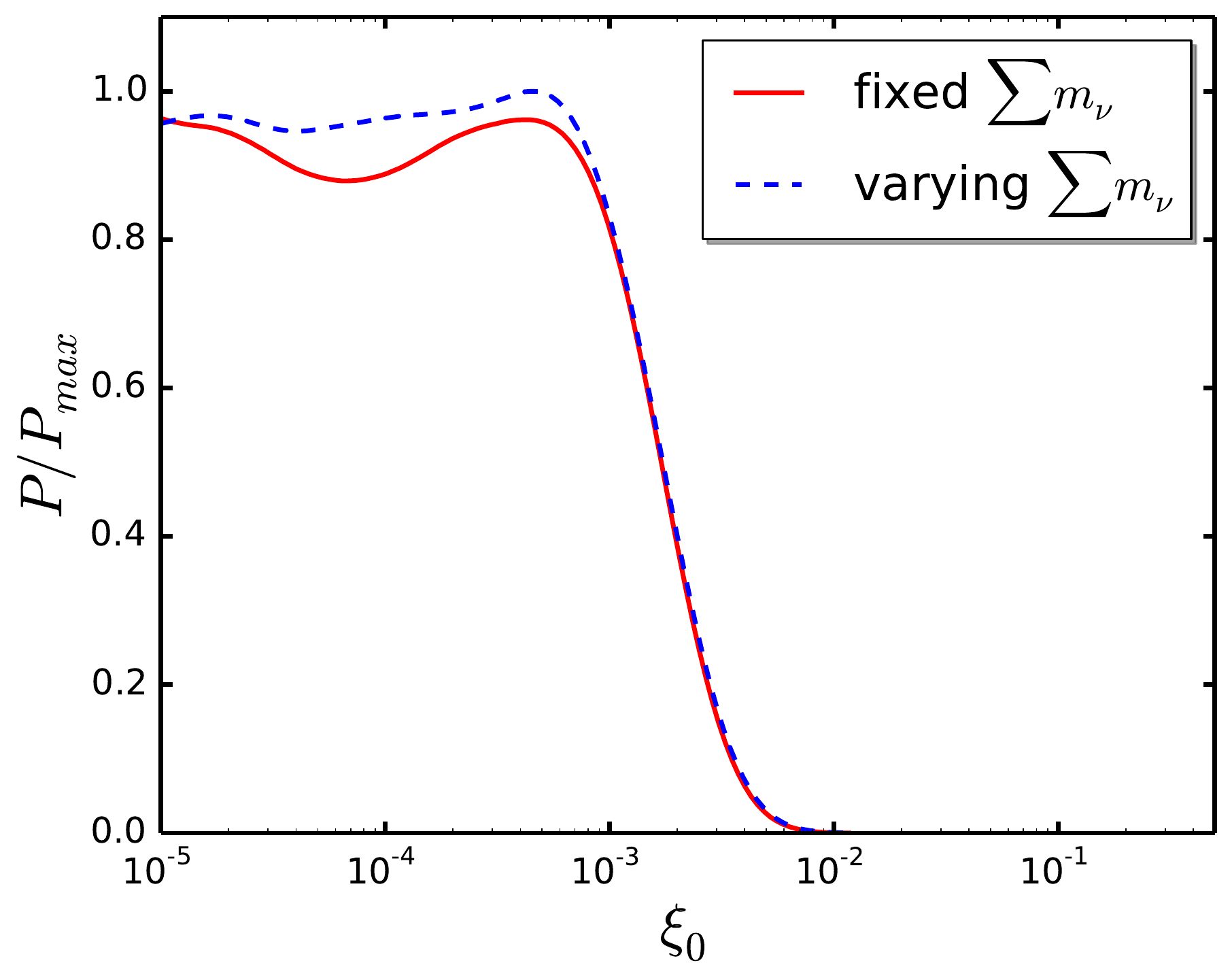}
\caption{\label{fig:dilaton1D}
Marginalized posterior distribution for $\xi_0$ in the Dilatons model with $\beta_0 = 5$. The datasets used in the analysis are PLC+BAO+lensing+MPK+WL as described in section~\ref{sec:current_data}. The red solid line shows the case with massive neutrinos with a fixed mass of $\sum m_{\nu} = 0.06$eV, while the blue solid lines shows the PDF after marginalizing over a varying $\sum m_{\nu}$. }
\end{center}
\end{figure}

\begin{table}[htp]
\centering
\begin{tabular}{c|c||c|c}
\cline{2-4}
 \multicolumn{1}{c|}{}& Fixed $\sum m_{\nu}$ & \multicolumn{2}{|c}{Varying $\sum m_{\nu}$} \\
\cline{2-4}
\cline{2-4}
\hline
\multirow{2}{*}{Symmetron} & $\xi_{\star}$ & $\xi_{\star}$ & $\sum m_{\nu}$ \\
\cline{2-4}
 & $0.8 \, (1.5) \times 10^{-3}$& $ 0.9 \,(1.8) \times 10^{-3}$ &  $ 0.16\,(0.27)$\\
\hline
\hline
\multirow{2}{*}{Dilaton} & $\xi_0$  & $\xi_0$  & $\sum m_{\nu}$ \\
\cline{2-4}
& $ 2.1 \, (3) \times 10^{-3}$ & $ 2.3 \, (3) \times 10^{-3}$ & $ 0.15 \, (0.25)$\\
\hline
\end{tabular}
\caption{\label{tab:summ_fit}Summary of the 95\% CL upper limits of the MG parameters and the sum of neutrino masses (in unit of eV) derived from current observations described in Sec.~\ref{sec:current_data}. }
\end{table}

Analogously to the Symmetron model, we constrain the inverse mass parameter $\xi_0$ defined by Eq.~(\ref{eq:xi0}), and fix $\beta_0$ to a constant. Fig.~\ref{fig:dilaton1D} shows the posterior distribution for $\xi_0$ with the current value of the coupling parameter fixed at $\beta_0=5$. We find an upper bound of $\xi_0 < 3 \times 10^{-3}$ (95 \% C.L.). As for symmetrons, the sensitivity to the coupling is weak due to the lack of data on linear scales. However, as we will see in Sec.~\ref{sec:forecasts}, constraints will improve significantly with future surveys. Our results for the Dilaton model are summarized in Table~\ref{tab:summ_fit}

\section{Forecasts}
\label{sec:forecasts}

Constraints on scalar gravitational interactions derived in the previous Section, using current information available on linear scales, are relatively weak when compared to bounds available from astrophysical tests. With improved redshift resolution, depth and sky coverage that future surveys will provide, the number of modes in the linear regime will dramatically increase. Thus, it is interesting to know if future constraints from linear scales can become compatible with astrophysical bounds. 

In what follows, we perform a series of Fisher forecasts for the model parameters described in the previous Section, using, where possible, the current bounds on model parameters as fiducial values in the forecast. Where there was no upper bound, we use fiducial values motivated by a combination of theoretical considerations and existing constraints from non-linear scales. We also perform a principal component analysis (PCA) of $m(a)$ for a fixed order unity coupling $\beta$, to see how well future datasets can constrain an evolving mass parameter.

\subsection{The data assumed in the forecast}

The data we consider in our forecast include CMB temperature anisotropy (T) and polarization (E) power spectra with characteristics of the Planck survey, weak lensing shear (WL) and galaxy number count (GC) from an LSST-like survey~\cite{LSST}, with the survey parameters adopted from \cite{Ivezic:2008fe}, and their cross-correlations. In some cases, we compare this to constraints expected from the Dark Energy Survey (DES) \cite{DES}.

Theoretical power spectra are calculated assuming the LSST (DES) GC data is partitioned into 10 (4) tomographic redshift bins, while the WL shear field is split into 6 (4) tomographic redshift bins. In addition, we assume a flat FRW geometry and vary $h$, 
$\Omega_ch^2$, $\Omega_bh^2$, $\tau$, $n_s$, $w$ and $A_s$, together with the modified gravity parameters. The fiducial values of the cosmological parameters are taken to be the Planck 2015 best fit results. To calculate the WL and GC auto- and cross-correlation spectra in our scalar-tensor models, we have applied the MGCAMB patch to CAMBSources~\cite{CAMBsources}.  The details of the implementation are described in \cite{Zhao:2008bn, Hojjati:2011xd}.

\subsection{Fisher analysis}

For a given model, one can calculate the Fisher matrix ~\cite{Fisher} to determine how well future surveys can constrain its parameters. The inverse of the Fisher matrix provides a lower bound on the covariance matrix of the model
parameters via the Cram$\acute{\rm e}$r-Rao inequality, ${\bf C} \geq {\bf F}^{-1}$. For zero-mean Gaussian-distributed
observables, such as the angular correlations $C^{XY}_\ell$, the Fisher matrix is given by
\be
F_{ab} =
f_{\rm sky} \sum_{\ell=\ell_{\rm min}}^{\ell_{\rm max}}\frac{2\ell +
1}{2} {\rm Tr}\left( \frac{\partial { C_\ell}}{\partial p_a} {
\tilde{C}_\ell^{-1}}\frac{\partial {C_\ell}}{\partial p_b} {
\tilde{C}_\ell^{-1}} \right) \ ,
\label{eq:Fisher}
\ee
where $p_{a}$ is the ${a}^{\rm th}$ parameter of our model and ${\bf
\tilde{C}_\ell}$ is the ``observed'' covariance matrix with elements
$\tilde{C}^{XY}_\ell$ that include contributions from noise:
\be
\tilde{C}^{XY}_\ell= C^{XY}_\ell+N^{XY}_\ell \ .
\label{eq:NoiseAdd}
\ee
 Eq.~(\ref{eq:Fisher}) assumes that all fields
$X(\hat{\bf n})$ are measured over contiguous regions covering a
fraction $f_{\rm sky}$ of the sky. The value of the lowest multipole
can be approximately inferred from $\ell_{\rm min} \approx \pi
/(2f_{\rm sky})$.  The noise matrix $N^{XY}_\ell$ includes the statistical noise as
well  as the  expected systematic errors. We refer the reader to \cite{Zhao:2008bn,Hojjati:2011xd} for the details of the Fisher matrix calculations for the individual experiments considered in our analysis. 

\subsection{The f(R) forecast}

\begin{figure}[tbp]
\includegraphics[width=0.47\textwidth]{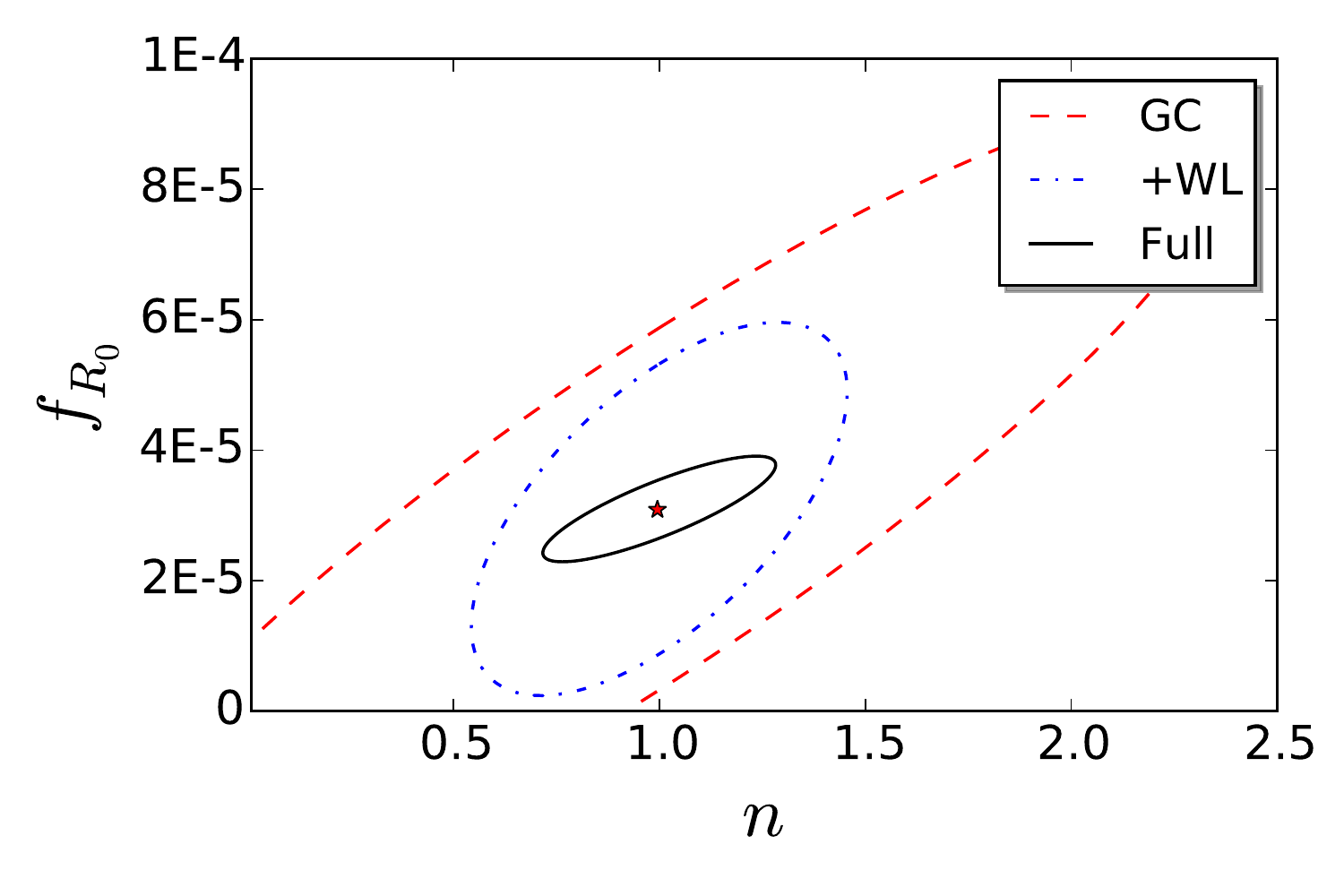}
\caption{\label{fig:fisher_husaw1}
Expected 1$\sigma$ bounds on the parameters of the Hu-Sawicki model. The assumed fiducial model is marked with a star. The importance of using the``Full'' set of observables (WL, GC and their cross-correlation) is clearly demonstrated. The Planck CMB data is included in all cases and is important for constraining the standard cosmological parameters.}
\end{figure}

In Fig.~\ref{fig:fisher_husaw1} we show 1$\sigma$ constraints on parameters of the Hu-Sawicki $f(R)$ model, as expected from LSST+ (LSST WL + LSST GC + Planck CMB). Recall that current data is unable to constrain $f_{R_0}$ unless one assumes a fixed value for $n$, since the two parameters are highly degenerate. Thus, the forecast in Fig.~\ref{fig:fisher_husaw1} depends strongly on the assumed fiducial value, indicated with a $\star$ on the plot. What we see is that for $n \sim 1$ or smaller, future data will be able to constrain both parameters simultaneously. 

Fig.~\ref{fig:fisher_husaw1} also shows the importance of including the cross-correlation between WL and GC. The information from GC alone is largely diluted by the unknown galaxy bias. Weak lensing, while not sensitive to the bias, is plagued by degeneracies coming from projection effects. Combining them helps determine the bias and break the degeneracies coming from projections. 

\begin{figure}[tbp]
\begin{center}
\includegraphics[width=0.45\textwidth]{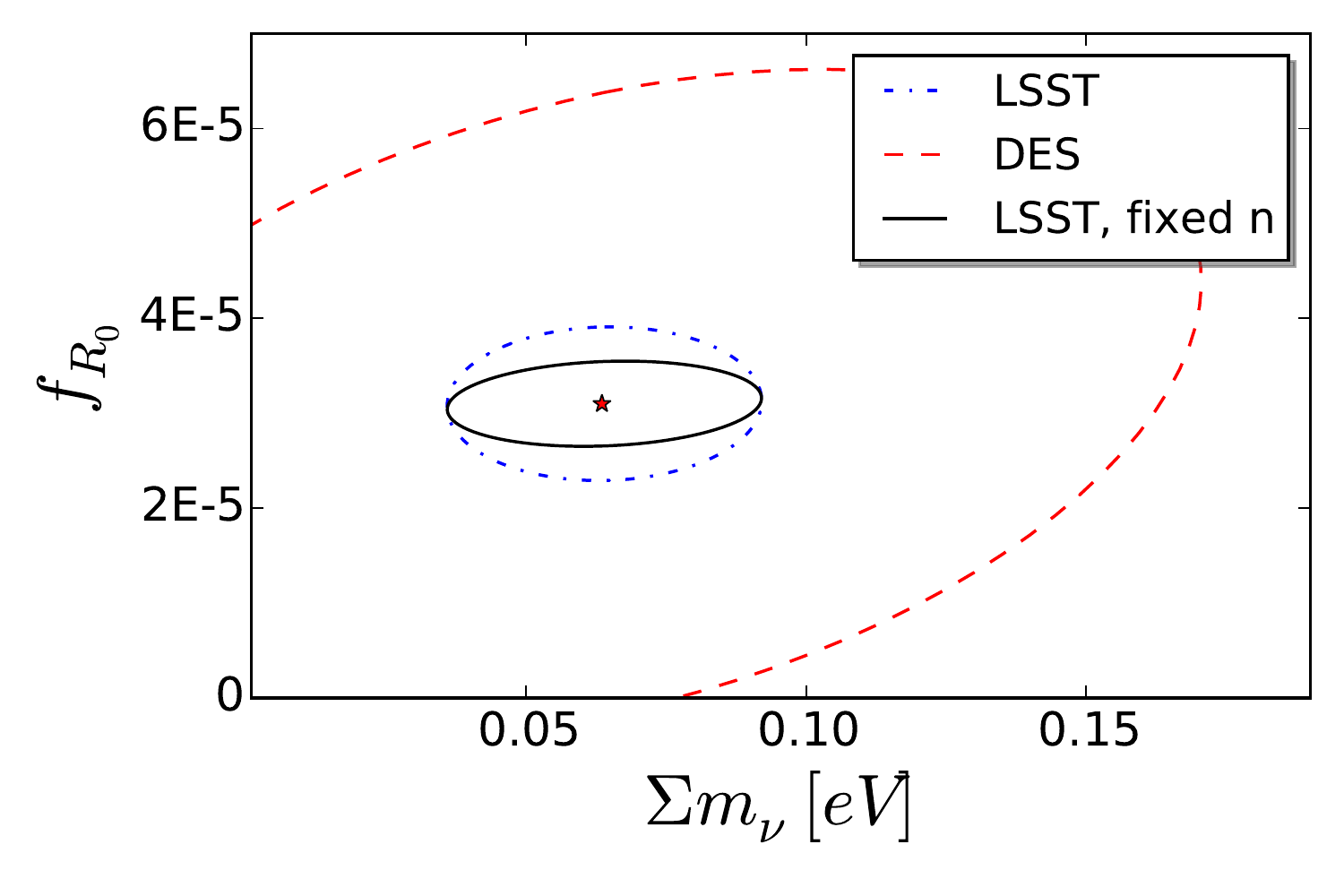}
\caption{
Comparison of the uncertainties expected from LSST+ vs those from DES+  for the $f_{R_0}$ paramater of the $n=1$ Hu-Sawicki model and the total mass of neutrinos. The assumed fiducial model is marked with a star. The effect of fixing $n$, as opposed to marginalizing over it, is also shown.}
\label{fig:fisher_husaw2}
\end{center}
\end{figure}

Fig.~\ref{fig:fisher_husaw2} compares joint 1$\sigma$ constraints on $f_{R_0}$ and the combined mass of neutrinos, $\sum m_\nu$, as expected from LSST+ vs those expected from DES+. We see that LSST+ can reduce uncertainties in both parameters by a factor of $3$. The plot shows the effect of marginalizing over $n$, however the outcome depends on the assumed fiducial value of $n$ (which is $n=1$).

\subsection{The Symmetron forecast}

\begin{figure}[tbp]
\includegraphics[width=0.5\textwidth]{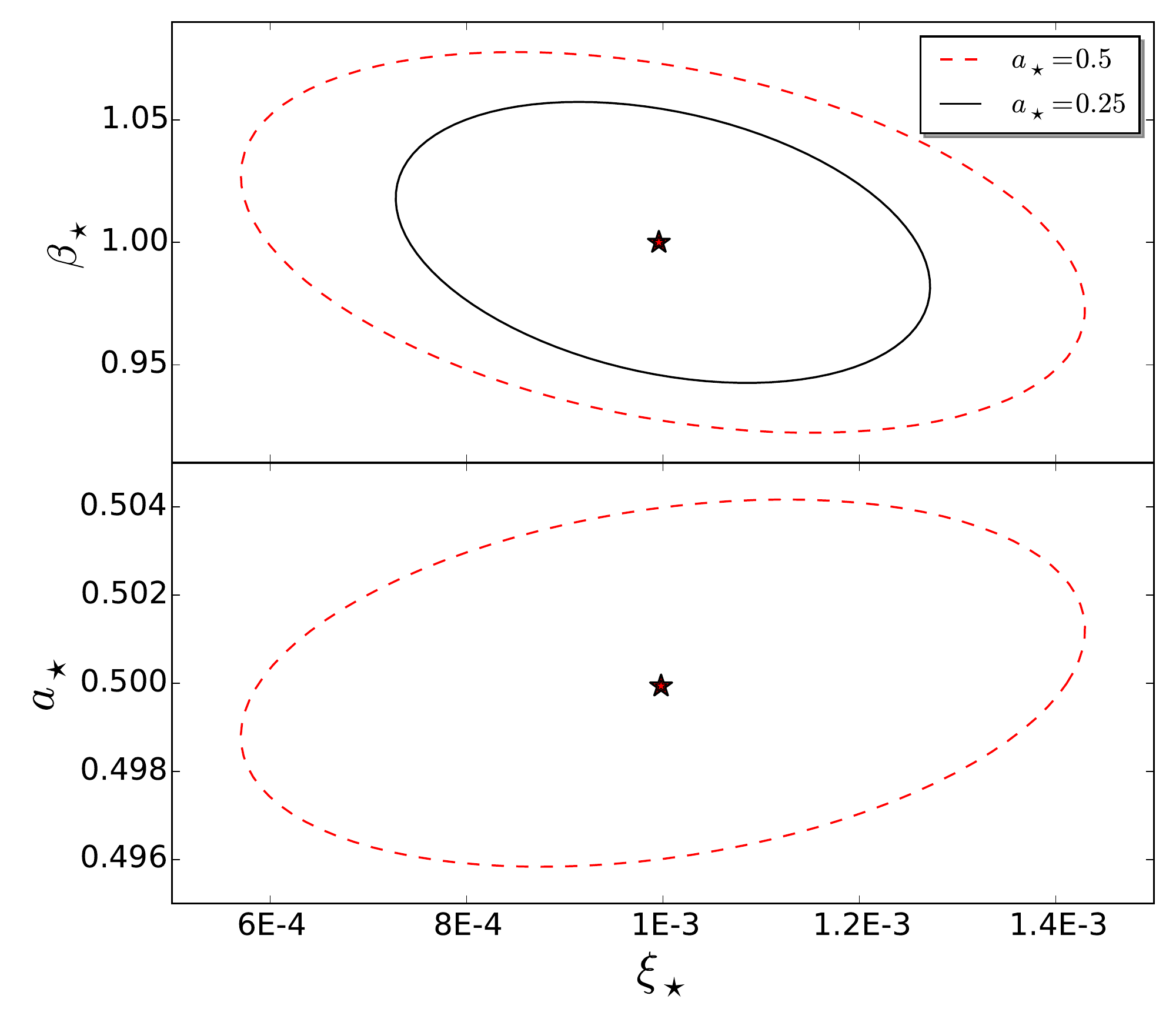}
\caption{\label{fig:fisher_symmetron}
Expected 1$\sigma$ constrains from LSST+ on the parameters of the Symmetron model. The assumed fiducial models are marked with a star. Unlike current data, LSST+ can simultaneously constrain $\beta_{\star} $ and $a_{\star}$ to a few percent level, and will improve the current bounds on $\xi_{\star}$. See Table~\ref{tab:comparison} for a quantitative comparison.}
\end{figure}
 
Fig.~\ref{fig:fisher_symmetron} shows the bounds on the parameters of the Symmetron parameters expected from LSST+. As a fiducial model, we assume $\beta_{\star} = 1$ and a mass scale of $\xi_{\star} = 10^{-3}$, which corresponds to a range of a few Mpc.  Current data is unable to constrain $\xi_{\star}$ if $a_{\star}=0.5$ or larger. For this reason, the bound on $\xi_{\star}$ in Sec.~\ref{sec:constraints} was derived for a fixed $a_{\star}=0.25$. We perform a forecast using two different fiducial values: $a_{\star} = 0.25$ and $0.5$. In the former case, LSST+  clearly improves on the bound in Sec.~\ref{sec:constraints}, even after marginalizing over $a_\star$ and $\beta_\star$. It will also be able to provide a non-trivial bound on $\xi_\star$ for $a_{\star} = 0.5$, which is the value assumed in much of the previous literature. The current and expected bounds are summarized in Table~\ref{tab:comparison}.

\begin{figure}[tbp]
\includegraphics[width=0.45\textwidth]{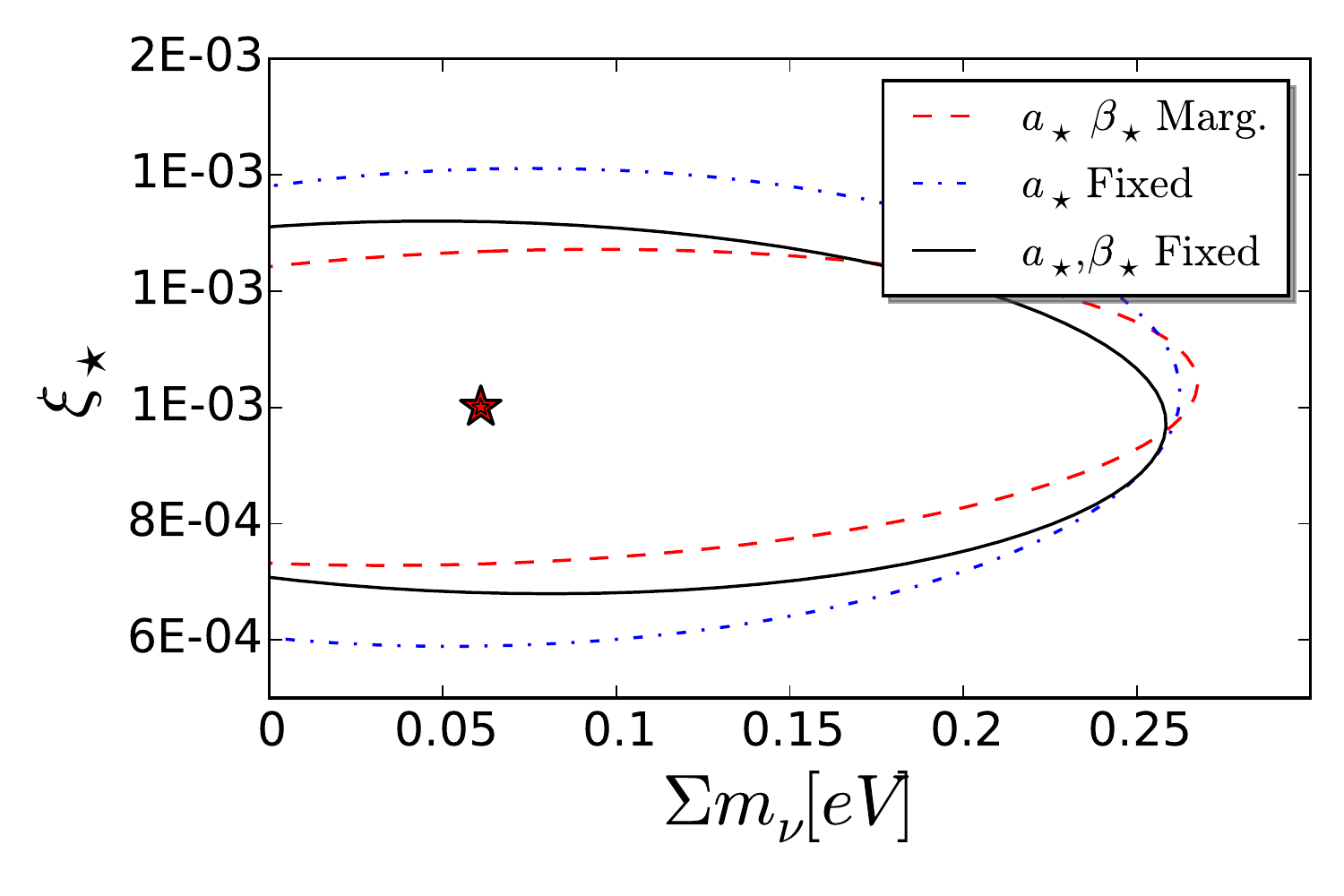}
\caption{\label{fig:fisher_symmetron_mnu}
Expected 1$\sigma$ bounds on the $\xi_{\star}$ parameter of the Symmetron model and the mass of neutrinos, $\sum m_{\nu}$. The assumed fiducial model is marked with a star. Fixing the other MG parameters in this model, as opposed to marginalizing over them, does not change the degree of degeneracy, neither it improves the constraints. }
\end{figure}

It is interesting to examine the possible degeneracy between the Symmetron parameters and the total mass of neutrinos. Fig.~\ref{fig:fisher_symmetron_mnu} shows the joint uncertainties in $\xi_\star$ and $\sum m_\nu$ expected from LSST+ assuming a fiducial model with $\beta_{\star} = 1$, $\xi_{\star} = 10^{-3}$, $a_{\star}=0.5$ and $\sum m_\nu=0.06$eV. It is clear from the figure that there is practically no degeneracy between $\xi_\star$ and $m_\nu$which is because they affect the growth on different scales. Fixing the other MG parameters in this model, as opposed to marginalizing over them, does not change the degree of degeneracy, neither it improves the constraints. 

\subsection{The Dilaton forecast}

\begin{figure}[tbp]
\includegraphics[width=0.45\textwidth]{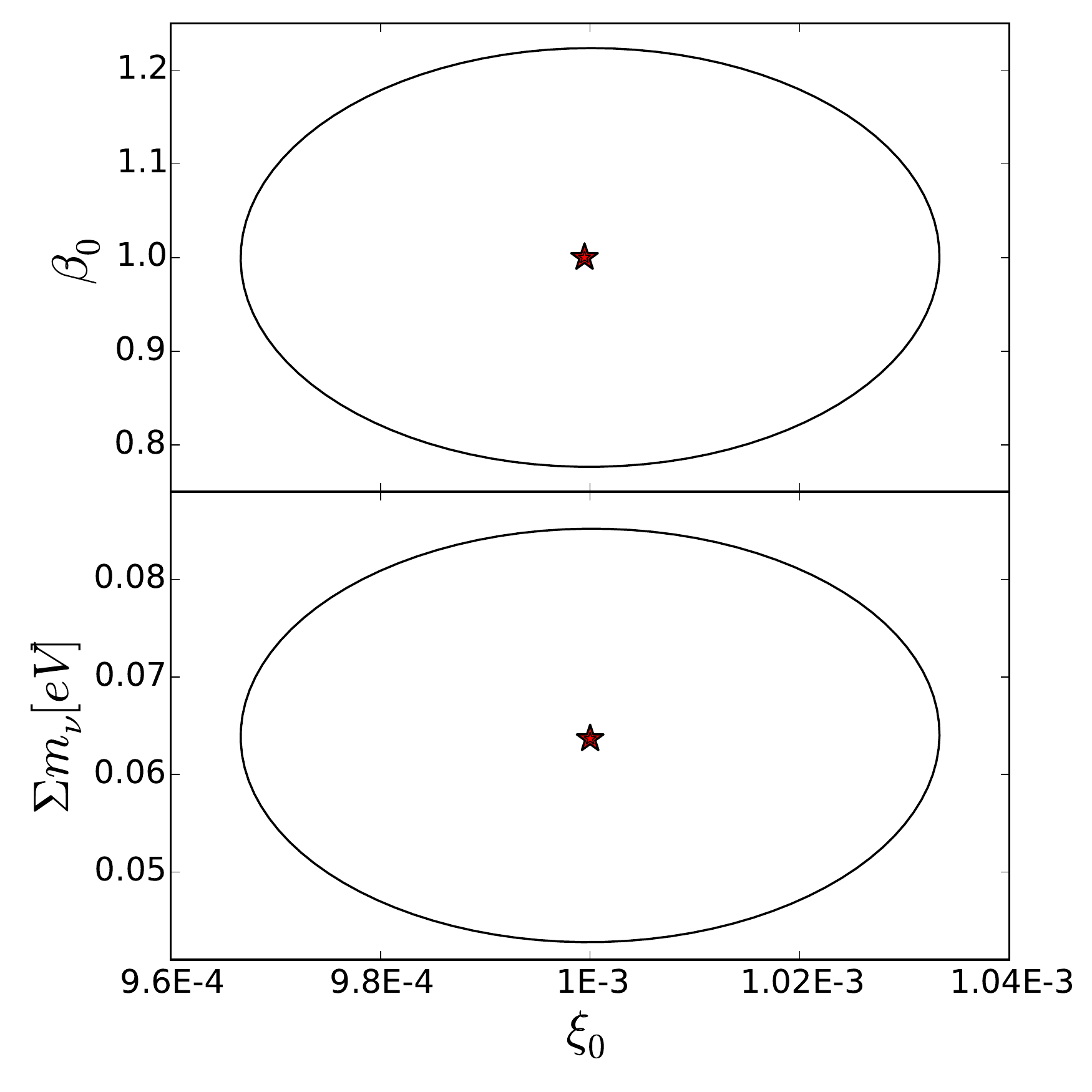}
\caption{\label{fig:fisher_dilaton}
1$\sigma$ bounds on the neutrino masses and parameters of the Dilaton model expected from LSST+. The fiducial values are marked with stars.}
\end{figure}

Fig.~\ref{fig:fisher_dilaton} shows expected bounds on the Dilaton model parameters, with $\beta_0=1$ and $\xi_0 = 10^{-3}$ as the fiducial values. Similar to the Symmetron case, we find that an LSST-like survey can constrain the inverse mass parameter $\xi_0$ to a percent level accuracy which is a significant improvement over current constraints. Constraints on the coupling constant $\beta_0$, however, are not as tight as those on $\beta_{\star}$ in the Symmetron case. This is due to a lesser impact of the Dilaton on the linear matter power spectrum. One can see from Fig.~\ref{fig:samples} that for the chosen fiducial values, $P(k)$ would deviate from the LCDM prediction far less in the Dilaton case compared to the Symmetron. The bottom panel in Fig.~\ref{fig:fisher_dilaton} shows the expected joint constraints on the neutrino masses, which are tighter than those for the Symmetron. Again, this is because Dilatons have a much lesser impact on the growth on linear scales.

\subsection{The Generalized Chameleon model}

\begin{figure}[tbp]
\includegraphics[width=0.5\textwidth]{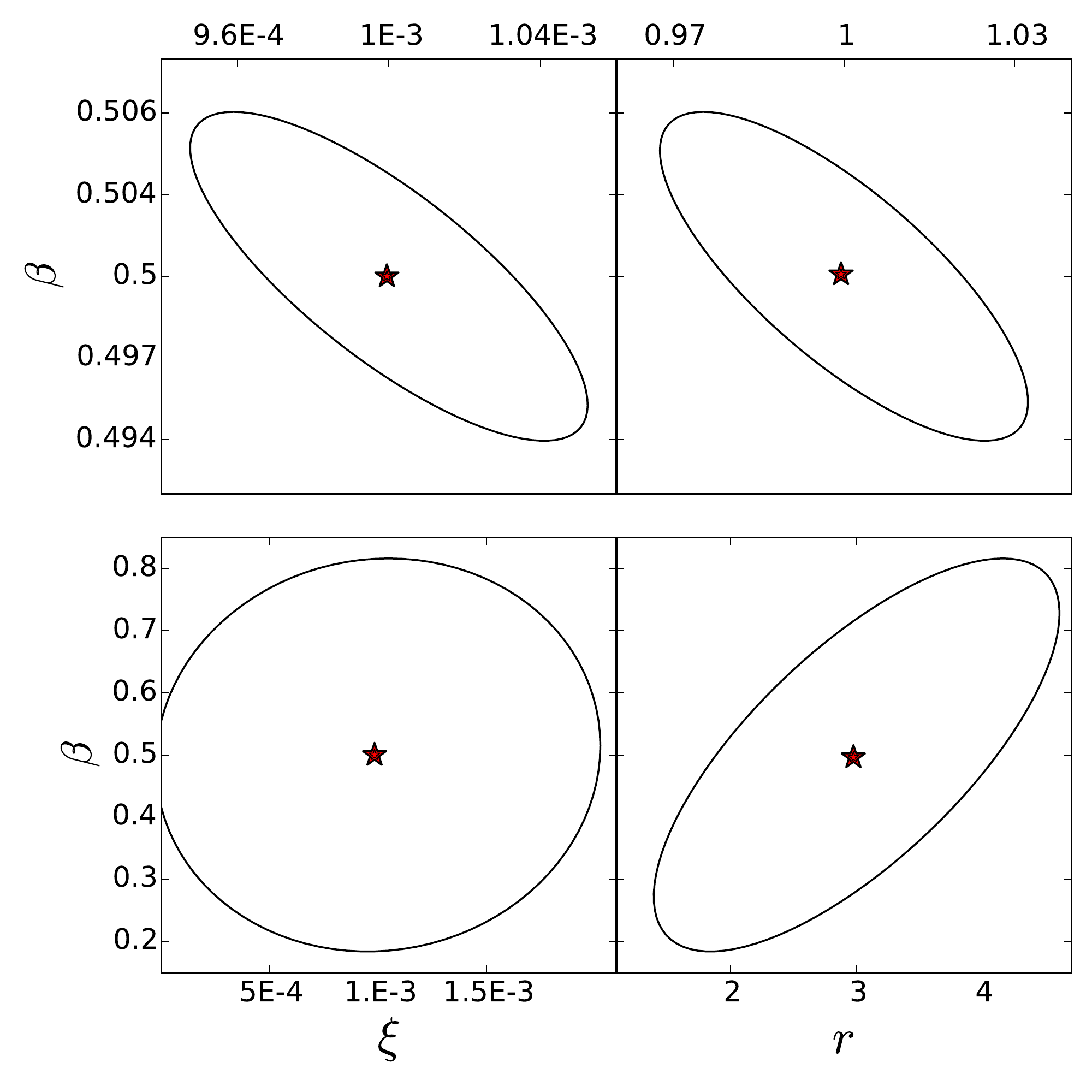}
\caption{\label{fig:fisher_gencham}
Expected 1$\sigma$ constrains on the the Generalized Chameleon parameters for a fiducial model with $r=3$ (top) and $r=1$ (bottom) as a fiducial model. In each case, the value of $r$ is varied and marginalized over.}
\end{figure}

Forecasts for the Generalized Chameleon provide a general estimate of how well one could constrain the scalar gravitational interactions with a next generation WL survey such as LSST. In Fig.~\ref{fig:fisher_gencham} we show forecasted uncertainties on the parameters of the generalized Chameleon model for two fiducial values of $r$, assuming that the coupling is constant ($s=0$). For a slower evolution with time ($r=1$), the scalaron mass decreases slower and modification to growth extends back to larger redshifts, leading to significantly tighter constraints. Thus, while LSST+ can constrain the coupling, the mass and the time-variation of the scalaron mass simultaneously, the strength of the bounds depends strongly on the assumed fiducial model.

\subsection{Principal Component Analysis of $m(a)$}
\label{PCA}

In addition to considering particular functional forms of $\beta(a)$ and $m(a)$ as motivated by the scalar-tensor models mentioned in preceding sections, it is also interesting to treat the coupling and the mass as two general functions and ask what features of these two functions can be constrained by the future data. In principle, one could discretize the functions $\beta(a)$ and $m(a)$ into $N$ bins in $a$ and treat the bins values as free parameters.  However, we find that even future data will not be able to simultaneously constrain $m(a)$ and $\beta(a)$ in a completely model-independent way, since the two parameters are largely degenerate in their effect on the observables on linear scales, as they appear together in $\epsilon(a,k)$ (see Eq.~(\ref{epsilon})). For this reason, we fix $\beta$ at a constant value of order unity and discretize $m(a)$ into bins with $m(a_i), i = 1,...,N$.  

As with earlier forecasts, we can calculate the Fisher matrix, and invert it to find the covariance matrix,
\be
\label{covariance}
C_{ij} \equiv \langle (p_i-{\bar p_i})(p_j-{\bar p_j})\rangle \ ,
\ee
where ${\bar p_i}$ are the ``fiducial'' values, and parameters include the bins $m(a_i)$, as well as the rest of cosmological parameters. We then isolate the $N\times N$ block of the matrix, $C^{m}$ corresponding to the covariance of $m(a_i)$ after marginalization over other parameters. Since the individual bins of $m(a_i)$ bins are highly correlated, the covariance matrix for these parameters will be non-diagonal, and the value of $m$ in any particular bin will be practically unconstrained. The Principal Component Analysis (PCA) \cite{Huterer:2002hy, Crittenden:2005wj,Zhao:2009fn,Hojjati:2011xd,Hojjati:2012ci}  is a way to decorrelate the parameters and find their linear combinations that are best constrained by data. Namely, we solve an eigenvalue problem to find a matrix $W^{m}$ that diagonalizes $C^{m}$:
\be
C^{m} = (W^{m})^T \Lambda W^{m} \ ;  \ \ \Lambda_{ij} = \lambda_i \delta_{ij} \ ,
\label{rotate}
\ee
where $W^{m}_{ij} \equiv \hat{e}_i(a_j)$ are the eigenvectors (or eigenmodes) and $\lambda_i$'s are the eigenvalues. In the limit of large $N$, one can write an arbitrary $m(a)$ as an expansion into $\hat{e}_i(a)$:
\be
m(a) - {\bar m(a)} = \sum_{i=1}^{N} \alpha_i \hat{e}_i(a)
\label{alphas}
\ee
in which case $\lambda_i$ can be interpreted as the variance of $\alpha_i$,
\be
\lambda_i = \sigma^2_{\alpha_i} \ .
\ee
It is customary to order the eigenmodes from the best constrained to the worst. Then the $i^{\rm th}$ eigenmode is referred to as the $i^{th}$ principal component (PC). Typically, one finds that only the first few modes are well constrained by the data, while most of them are practically unconstrained.

For our forecast, we partition $m(a)$ into $11$ bins, with $10$ of them evenly spaced in redshift within $z \in [0,3]$, and the $11$th bin ranging from $z = 3$ to $z = 30$. The last bin can be taken to be wide because the observables we work with are weakly sensitive to modifications at high redshifts. In what follows, we marginalize over the $11$th bin, since it is largely degenerate with some of the cosmological parameters, most prominently with $\Omega_m$. We take the fiducial model to be $\beta = 0.4$ and $m(a_i) = m_0$ for all $i$, with $m_0 = H_0/ \xi c$ and $\xi =10^{-3}$, corresponding to $m_0 = 0.2$ h/Mpc. 

\begin{figure*}[tbp]
\begin{center}
\includegraphics[width=0.45\textwidth]{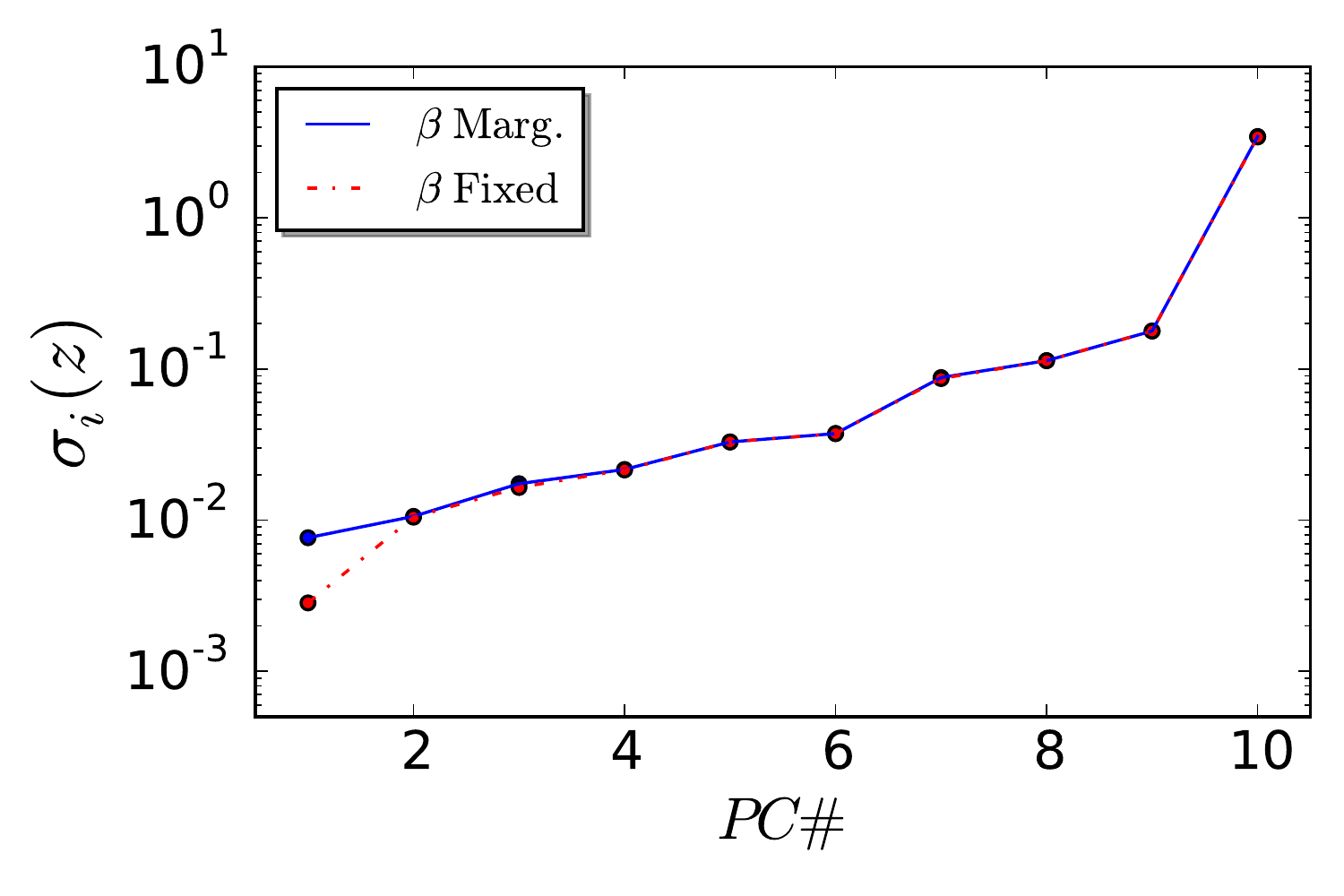}
\includegraphics[width=0.45\textwidth]{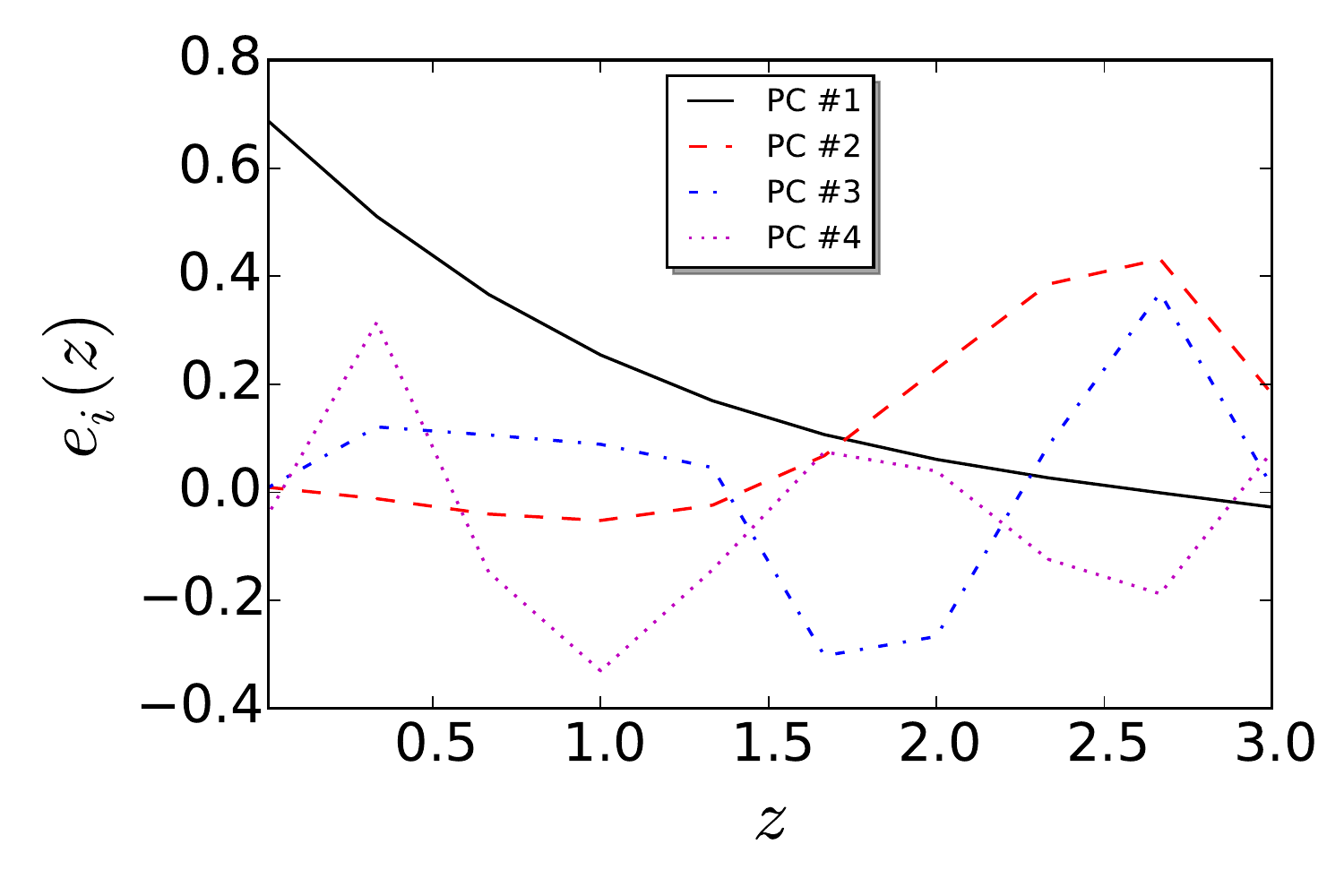}
\caption{\label{fig:pca_eigenvectors}
\emph{Left:} The uncertainties (square roots of eigenvalues) associated with the eigenmodes of $m(a)$ for the case when the coupling is fixed at $\beta = 0.4$ (solid line), and when it is marginalized over (dashed line). \emph{Right:} The first four best constrained eigenmodes of $m(a)$ after marginalizing over $\beta$. }
\end{center}
\end{figure*}

The left panel in Fig.~\ref{fig:pca_eigenvectors} shows the the forecasted uncertainties in the measurement of the eigenvectors for two cases: when $\beta$ is fixed, and when $\beta$ is marginalized over. In both cases, we marginalize over all cosmological parameters and the $11$th $m$-bin. The right panel in Fig.~\ref{fig:pca_eigenvectors} presents the first four best constrained eigenvectors of $m(a)$ after marginalizing over $\beta$. One can interpret the best constrained mode (PC1) as that corresponding to a weighted average value of $m(a)$. The second best constrained mode (PC2) has a single node and corresponds to the difference between the high-z and low-z values of $m(a)$. The third best mode (PC3) has 3 nodes, PC4 has 4 nodes,  and so on.

The eigenvalue plot demonstrates that marginalizing over $\beta$ affects the first eigenmode of $m(a)$, but not the others. This is because the main effect of a constant $\beta$ is an overall rescaling of the strength of the 5th force. It is largely degenerate with the average value of $m(a)$, but has no impact on the detectability of time-variation of $m(a)$. After marginalizing over $\beta$, LSST+ can measure one mass parameters (the average $m(a)$) with an accuracy that is better than $0.01$ h/Mpc, or about 5\% of the fiducial $m_0$, and another $3$ parameters, describing more rapid evolution of the mass with time, with accuracy better than $0.02$ h/Mpc, or 10\% of the fiducial value. 

The extrema of the eigenmodes indicate the ``sweet spots'' in redshift, or epochs at which variations in $m(a)$ are best constrained with LSST+. It is evident from the right panel in Fig.~\ref{fig:pca_eigenvectors}, for instance the shape of PC2, that LSST+ is more sensitive to time-variations at $z>1.5$. This is because at higher redshifts there is a larger number of Fourier modes that are still in the linear regime.

\section{Discussion and Conclusions}

\begin{table*}[htp]
\centering
\begin{tabular}{c|c|c||c|c|c||c|c|}
\cline{2-8}
& \multicolumn{2}{|c||}{Hu-Sawicki $f(R)$} & \multicolumn{3}{|c||}{Symmetron} & \multicolumn{2}{|c|}{Dilaton} \\
\cline{2-8}
 Parameters & $f_{R_0}$ & $n=1$ & $\xi_{\star}$ & $\beta_{\star}=1$ & $a_{\star}=0.25 \ (0.5)$ & $\xi_0$ & $\beta_0=1 \ (5)$ \\
\hline
Current 1$\sigma$ & $4 \times 10^{-5}$ & - & $10^{-3}$ & - & - & {\rm unconstrained} ($2.3 \times 10^{-3}$) & - \\  
\hline
LSST+ 1$\sigma$   &   $6 \times 10^{-6}$  &  $0.3$   &   $2 \ (2.9) \times 10^{-4}$   &   $0.05 \ (0.07)$  &   $0.001 \ (0.005)$    &     $2.7 \times 10^{-5}$   &   $2.3 \times 10^{-1}$     \\
\hline
\end{tabular}
\caption{\label{tab:comparison}
The current 68 \% C.L. uncertainties and those expected from LSST+. The blocks with ``$-$'' mean the parameter was fixed at its fiducial value. The values in parenthesis indicate those obtained for an alternative fiducial value.  }
\end{table*}

Modifications of gravity on cosmological scales can potentially explain the origin of cosmic acceleration. The Generalized Brans-Dicke theory, in which there is an additional scalar degree of freedom that mediates a fifth force, is one of the viable MG models that are able to fit observations after the required tuning of model parameters. 

In this work, we have investigated the observational constraints on three MG models within the general framework of the GBD theory, namely, the $f(R)$, the Symmetron and the Dilaton models, using latest observations of CMB, BAO, weak lensing and galaxy clustering. In all cases, we used observables on linear scales to avoid the complexities of the modelling of nonlinearities and redshift-space distortions. 

We find that the $\Lambda$CDM model is consistent with all observations. Specifically, we find the constraint on $f_{R_0}$, the model parameter in the Hu-Sawicki $f(R)$ model, to be $f_{R_0}<8\times10^{-5}$ (95\% CL) when the sum of neutrino masses is fixed to be 0.06 eV. Since both massive neutrinos and MG models studied in this paper can alter the structure growth in a scale-dependent way, a degeneracy is expected. Therefore we perform another analysis with the neutrino mass varying, and we find that the constraint is diluted to $f_{R_0}<1.0\times10^{-4}$ (95\% CL). For the Symmetron model, the 95\% CL upper limit is $\xi_{\star}< 1.8\times10^{-3}$ with $\beta_\star$ and $a_\star$ fixed at $1$ and $0.25$, respectively. For the Dilaton model, we find $\xi_{0}<3\times10^{-3}$ at 95\% CL when $\beta_0=5$. Tables \ref{tab:husaw_AL} and \ref{tab:summ_fit} summarize the current bounds.

We have also performed a forecast for ongoing and upcoming imaging surveys including DES and LSST, and present the results in Sec. \ref{sec:forecasts}.  A comparison between the current and future constraints on model parameters is shown in Table \ref{tab:comparison}. As one can see, the improvement is significant and, despite the high level of degeneracy, more than one parameter can be constrained simultaneously. In the Hu-Sawicki model, the upper limit of $f_{R_0}$ is reduced by a factor of $6.7$ and $n$ can be constrained with $\approx 25\%$ accuracy for $n=1$. For the Dilaton model, current data is unable to constrain $\xi_0$ if $\beta_0=1$. However, we find that LSST+ can simultaneously constrain $\xi_{0}$ at $\sim {\rm few} \times10^{-5}$  and measure $\beta_0 \sim 1$ with $\approx 20\%$ accuracy. In the Symmetron model, the constraint on $\xi_{\star}$ is  improved by a factor of $3$, while simultaneously constraining $a_{\star}$ and $\beta_{\star}$ within a few percent. This is compatible with current bounds derived from astrophysical tests, such as the cluster profile \cite{MGcluster}, galactic dynamics and so on, which requires high-resolution N-body or hydrodynamical simulations \cite{Nbody1,Nbody2} of the MG models. Additionally, to demonstrate the capabilities of an LSST-like survey, we have presented constraints on the Generalized Chameleon model in Fig.~\ref{fig:fisher_gencham}.

Given the power of future surveys, a model-independent analysis will become possible. In this work, we performed a PCA study of $m(a)$, to forecast the maximum number of parameters of the scalaron mass function that can be well determined. We find that an LSST-like survey will be able to measure the average mass parameter with an accuracy of $0.01$ h/Mpc and another $3$ parameters quantifying the time-variation of $m(a)$ with an accuracy that is better than $0.02$ h/Mpc. Finally, we note that future spectroscopic and HI surveys, such as eBOSS and SKA \cite{eBOSS,SKA}, will also provide powerful constraints on MG parameters that will be highly complementary to those from a photometric survey like LSST \cite{LSSTSKA}.

\acknowledgements

The research of AH, AP, AZ and LP is supported by the Natural Sciences and Engineering Research Council of Canada (NSERC). AP was sponsored in part by the Robert Frazier Memorial Fellowship. AZ is supported in part by the Bert Henry Memorial Entrance Scholarship at SFU. GBZ is supported by the Strategic Priority Research Program "The Emergence of Cosmological Structures" of the Chinese Academy of Sciences Grant No. XDB09000000. This research was enabled in part by support provided by WestGrid \cite{westgrid} and Compute Canada \cite{compute}. P.B. acknowledges partial support from the European Union FP7 ITN INVISIBLES (Marie Curie Actions, PITN- GA-2011- 289442) and from the Agence Nationale de la Recherche under contract ANR 2010 BLANC 0413 01. ACD acknowledges partial support from STFC under grants ST/L000385/1 and ST/L000636/1.

\end{document}